\documentclass[twocolumn,prl,floatfix,preprintnumbers,a4paper,nofootinbib,superscriptaddress]{revtex4-1}
\usepackage{hyperref}
\usepackage{amsmath,amsfonts}
\usepackage{graphicx}
\usepackage{color}
\usepackage{soul}
\usepackage{ifpdf}
\usepackage{ulem}   
\usepackage{enumerate}
\usepackage{multirow}
\usepackage{acronym}
\usepackage{float}
\definecolor{LinkColor}{rgb}{0.75, 0, 0}
\definecolor{CiteColor}{rgb}{0, 0.5, 0.5}
\definecolor{UrlColor}{rgb}{0, 0, 0.75}
\hypersetup{linkcolor=LinkColor}
\hypersetup{citecolor=CiteColor}
\hypersetup{urlcolor=UrlColor}
\usepackage[utf8]{inputenc}
\usepackage{ulem}
\newcommand\xquote[1]{``#1"}
\newcommand\psinr[2]{$\psi_{{#1}{#2}}^{\mathrm{NR}}$}

\def\fig#1{Fig. #1}
\def\eqn#1{Eq.(#1)}
\def\bh#1{black hole#1
  (BH#1)\gdef\bh{BH}}
\def\et#1{Einstein Telescope#1
  (ET#1)\gdef\et{ET}}
\def\aligo#1{Advanced LIGO#1
  (Adv. LIGO#1)\gdef\aligo{Adv. LIGO}}
\def\snr#1{signal to noise ratio#1
  (SNR#1)\gdef\snr{SNR}}
\def\bbh#1{binary black hole#1
  (BBH#1)\gdef\bbh{BBH}}
\def\qnm#1{Quasi-Normal Mode#1
  (QNM#1)\gdef\qnm{QNM}}
\def\gw#1{gravitational wave#1
  (GW#1)\gdef\gw{GW}}
\def\pn#1{post-Newtonian#1
  (PN#1)\gdef\pn{PN}}
\def\pnl#1{post-Newtonian-like#1
  (PN-like#1)\gdef\pnl{PN-like}}
\def\nr{numerical relativity
  (NR)\gdef\nr{NR}}
\def\pt{\bh{} perturbation theory\gdef\pt{perturbation theory}}
\def\GOLS#1{\textit{greedy-OLS}#1}
\def\rd{ringdown}
\def\bbc#1{binary black hole coalescence#1
  (BBC#1)\gdef\bbc{BBC}}

\newcommand*{\factor}{0.5}

\definecolor{lightblue}{rgb}{.82,.88,0.95}
\definecolor{lightred}{rgb}{0.95,.86,0.86}
\definecolor{yellow}{rgb}{0.95,0.95,0.86}
\definecolor{green}{rgb}{.90,1,0.95}
\definecolor{lightpurple}{rgb}{.95,0.85,0.95}

\newcommand{\eqns}[2]{Equations~(\ref{#1}-\ref{#2})}

\begin{document}

\title{ {Modeling ringdown: Beyond the fundamental quasinormal modes } }
\author{Lionel London}
\affiliation{Center for Relativistic Astrophysics, School of Physics,
 Georgia Institute of Technology, Atlanta, Georgia 30332, USA}
 \author{James Healy}
\affiliation{Center for Computational Relativity and Gravitation,
School of Mathematical Sciences, Rochester Institute of Technology, Rochester, New York 14623, USA}
 \author{Deirdre Shoemaker}
 \affiliation{Center for Relativistic Astrophysics, School of Physics,
 Georgia Institute of Technology, Atlanta, Georgia 30332, USA}

\begin{abstract}
While black hole perturbation theory predicts a rich quasi-normal mode structure, technical challenges have limited the numerical study of  excitations to the fundamental, lowest order modes caused by the coalescence of black holes.
Here, we present a robust method to identify quasinormal mode excitations beyond the fundamentals within currently available numerical relativity waveforms.
In applying this method to waveforms of 68 initially nonspinning black hole binaries, of mass-ratios 1:1 to 15:1, we find not only the fundamental quasinormal mode amplitudes, but also overtones, and evidence for second order quasinormal modes.
We find that the mass-ratio dependence of quasinormal mode excitation is very well modeled by a post-Newtonian-like sum in symmetric mass-ratio.
Concurrently, we find that the mass-ratio dependence of some quasinormal mode excitations is qualitatively different from their post-Newtonian inspired counterparts, suggesting that the imprints of nonlinear merger are more evident in some modes than in others.
We present new fitting formulas for the related quasinormal mode excitations, as well as for remnant black hole spin and mass.
We also discuss the relevance of our results in terms of gravitational wave detection and characterization.
\end{abstract}
\maketitle%
%
\section{Introduction}
\label{sec:intro}
%
\par As we approach the era of \gw{} detection, there is a tremendous effort to understand and predict the rich gravitational wave signals coming from all expected sources of radiation.
These predictions are used to construct \gw{} templates that will enable not only the recognition of \gw{} signals within noise, but also the extraction of information about the source.
It is for these purposes that the development of templates that include the final moments of \bbh{} coalescence is important for future \gw{} detection.
\par While source populations remain uncertain, \bbh{} systems are expected to account for multiple signals per year and, if systems with a total mass of a few hundred times that of our sun or larger are observed, detectors such as Advanced LIGO and the Einstein Telescope are most sensitive to the final stages of \bbh{} coalescence \cite{Abadie:2010cf,Belczynski:2010tb,Mishra:2010tp,Ajith:2009fz}.
In these final moments the two \bh{s} merge into a perturbed, remnant \bh{}, whose gravitational radiation \textit{rings down} like a struck bell.
Very roughly put, if one were to observe the remnant at an orientation $(\theta,\phi)$ relative to its spin axis, and at a distance $r$ away, then the observable time domain strain of this decaying \textit{\rd{}} radiation may be written as the real part of
\begin{align} \label{eq:example_sum}
h \; &= \; -\frac{1}{r} \; \sum_{l,m,n} A_{lmn}\,S_{lmn}(\theta,\phi)\,\frac{e^{i(\omega_{lmn}+i/\tau_{lmn})\,t}}{(\omega_{lmn}+i/\tau_{lmn})^2}
\\ \nonumber
 &= \; h_+ \, - \, i \, h_\times \, \;.
\end{align}
Here, $h_+$ and $h_\times$ are the real valued plus and cross polarization states. In general, a linear combination of these states will be detected \cite{Alcubierre.0707.4654,2014arXiv1410.8310T}.
\par If provided the remnant \bh{'s} mass and spin, then the perturbation theory of isolated Kerr \bh{s} informs us of \eqn{\ref{eq:example_sum}}'s spatial multipoles and temporal frequencies: the \qnm{s} that dominate \rd{} \cite{PreTeu73_2,leaver85,PhysRevD76Berti}.
However, in order to model astrophysically relevant \rd{} signals the output of \nr{} simulation is generally needed to tell us how much each multipolar component is excited for a given initial binary \cite{PhysRevD76Berti,Kamaretsos:2011um}.
%
\begin{figure}[t]
\includegraphics[width=\factor\textwidth]{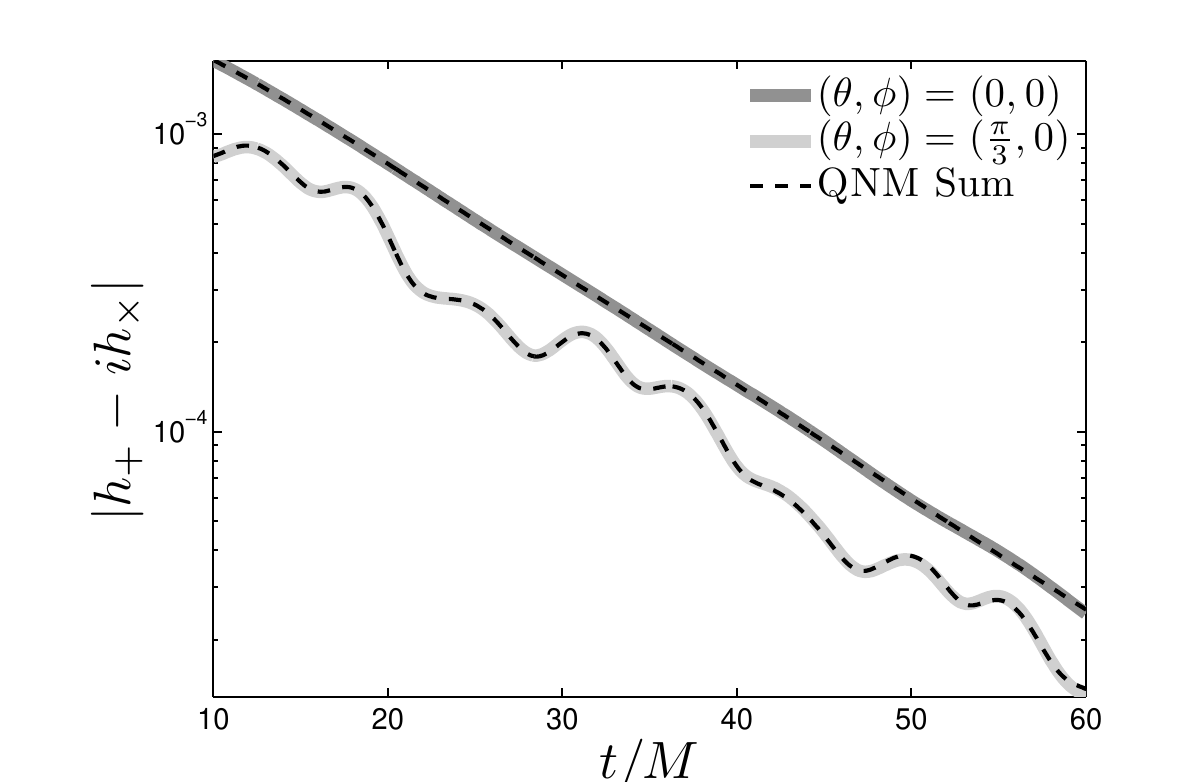}
\caption{Ringdown for a 2:1 mass-ratio, initially nonspinning \bh{} binary calculated via the GaTech \texttt{MAYA} code \cite{maya-note,et-web,cactus-web,Schnetter-etal-03b,Husa:2004ip}. The solid gray lines show the time domain envelope of \nr{} \rd{} for two different lines of sight. Here $\theta$ and $\phi$ are polar and azimuthal angles relative to the \bh{s} final spin vector. The dashed black lines show the corresponding model \rd{s} (\qnm{} sums) calculated using the results of this paper: estimation of spheroidal \qnm{} excitations from \nr{}, including and beyond the fundamental overtones.}
\label{fig:full_ringdown}
\end{figure}
%
\par For this reason, applying \pt{} to the analysis of \nr{} \rd{} has assisted in the creation of inspiral-merger-\rd{} templates \cite{Taracchini:2013rva,Damour:2009kr,Pan:2009wj}, and revealed novel relationships between the initial binary's configuration and the remnant \bh{'s} parameters \cite{Kamaretsos:2011um}.
But thus far, technical challenges have limited analysis primarily to the fundamental (lowest overtone) \qnm{s}, while it has also been acknowledged that a more detailed application of \pt{} to \nr{} \rd{} may be needed \cite{Kamaretsos:2011um,Taracchini:2013rva,Kelly:2012nd,Dorband:2006gg,Okuzumi:2008ej,Schnittman:2007ij,Buonanno:2006ui}.
As an example of \rd{}'s potential complexity, \fig{\ref{fig:full_ringdown}} shows the time domain strain envelope of a potential 2:1 mass-ratio \rd{} signal of an initially nonspinning \bh{} binary, observed at two different lines of sight.
Here we see that the sum of many \qnm{s} precisely models \nr{} \rd{} data.
This example case demonstrates that both the intrinsic \qnm{s} of \pt{} and the observer's extrinsic line of sight contribute to the richness of possible \rd{} signals.
\par In this study we assist in clarifying the extent to which \qnm{s} beyond the fundamentals are pertinent to the physics and modeling of \nr{} \rd{} (e.g.~\fig{\ref{fig:full_ringdown}}).
We consider the \rd{} of 68 initially nonspinning \bbh{} simulations of mass-ratios between 1:1 and 1:15.
In doing so, we find that \qnm{} excitation is exceptionally well modeled by a \pnl{} expansion (Sec.\ref{sec:MAPPING}).
However, we also find that the excitation amplitudes of some \qnm{s} differ qualitatively from their \pn{} counterparts, suggesting that the imprints of nonlinear merger are more evident in these \qnm{s} than in others (Sec.\ref{sec:DISCUSS:PTCOMMENTS}).
But first, we present a robust method to estimate multiple \qnm{s} within \nr{} \rd{} (Sec.\ref{sec:MULTI_METHOD}).
We then apply this method to a series of initially nonspinning \nr{} runs of varying mass-ratio (Sec.\ref{sec:MULTI_FITs}-\ref{sec:Beyond}).
Lastly, we consider the results of our analysis (overtones and second order modes) in the context of \rd{}-only templates (Sec.\ref{sec:DISCUSS:TEMPL}).
Generally, our results may be of use for the construction of merger-\rd{} templates.
\par A complete paper outline is given in Sec.~\ref{sec:intro_outline}.
A full summary of fitting formulas and coefficients for \qnm{} excitations is given in Appendix \ref{app:Fit_Coeffs}.
For convenience, fits for the most dominant \qnm{} excitation amplitudes in \eqn{\ref{eq:example_sum}} are below:
\begin{widetext}
\begin{align}
	\label{eq:fit_fun_A220}
	A_{220}(\eta) \; &= \; \tilde{\omega}_{220}^2  \, (\, 0.9252 \, e^{ 0.0000 i} \eta + 0.1323 \, e^{ 0.0000 i} \eta^{2} \, )\\
	\label{eq:fit_fun_A221}
	A_{221}(\eta) \; &= \; \tilde{\omega}_{221}^2  \, (\, 0.1275 \, e^{ 5.3106 i} \eta + 1.1882 \, e^{ 0.4873 i} \eta^{2} + 8.2709 \, e^{ 3.3895 i} \eta^{3} + 26.2329 \, e^{ 0.1372 i} \eta^{4} \, )\\
	\label{eq:fit_fun_A210}
	A_{210}(\eta) \; &= \; \tilde{\omega}_{210}^2 \, \sqrt{1-4\eta} \, (\, 0.4795 \, e^{ 3.5587 i} \eta + 1.1736 \, e^{ 1.5679 i} \eta^{2} + 1.2303 \, e^{ 6.0496 i} \eta^{3} \, )\\
	\label{eq:fit_fun_A330}
	A_{330}(\eta) \; &= \; \tilde{\omega}_{330}^2 \, \sqrt{1-4\eta} \, (\, 0.4247 \, e^{ 5.4979 i} \eta + 1.4742 \, e^{ 3.6524 i} \eta^{2} + 4.3139 \, e^{ 6.0787 i} \eta^{3} + 15.7264 \, e^{ 3.2053 i} \eta^{4} \, )\\
	\label{eq:fit_fun_A331}
	A_{331}(\eta) \; &= \; \tilde{\omega}_{331}^2 \, \sqrt{1-4\eta} \, (\, 0.1480 \, e^{ 2.9908 i} \eta + 1.4874 \, e^{ 0.5635 i} \eta^{2} + 10.1637 \, e^{ 4.2348 i} \eta^{3} + 29.4786 \, e^{ 1.7619 i} \eta^{4} \, )\\
	\label{eq:fit_fun_A320}
	A_{320}(\eta) \; &= \; \tilde{\omega}_{320}^2  \, (\, 0.1957 \, e^{ 5.8008 i} \eta + 1.5830 \, e^{ 3.2194 i} \eta^{2} + 5.0338 \, e^{ 0.6843 i} \eta^{3} + 3.7366 \, e^{ 4.1217 i} \eta^{4} \, )\\
	\label{eq:fit_fun_A440}
	A_{440}(\eta) \; &= \; \tilde{\omega}_{440}^2  \, (\, 0.2531 \, e^{ 1.5961 i} \eta + 2.4040 \, e^{ 5.1851 i} \eta^{2} + 14.7273 \, e^{ 1.9953 i} \eta^{3} + 67.3624 \, e^{ 4.9143 i} \eta^{4} + 126.5858 \, e^{ 1.8502 i} \eta^{5} \, )\\ 
	\label{eq:fit_fun_A430}
	A_{430}(\eta) \; &= \; \tilde{\omega}_{430}^2 \, \sqrt{1-4\eta} \, (\, 0.0938 \, e^{ 3.2607 i} \eta + 0.8273 \, e^{ 0.7704 i} \eta^{2} + 3.3385 \, e^{ 4.8264 i} \eta^{3} + 4.6639 \, e^{ 2.7047 i} \eta^{4} \, )\\
	\label{eq:fit_fun_A550}
	A_{550}(\eta) \; &= \; \tilde{\omega}_{550}^2 \, \sqrt{1-4\eta} \, (\, 0.1548 \, e^{ 5.3772 i} \eta + 1.5091 \, e^{ 2.5764 i} \eta^{2} + 8.9333 \, e^{ 5.5995 i} \eta^{3} + 42.3431 \, e^{ 2.1269 i} \eta^{4} + 89.1947 \, e^{ 5.3348 i} \eta^{5} \, )
\end{align}

\end{widetext}
%
Here, $M$ is the sum of the initial \bh{} masses, $$M = \mathrm{m_1+m_2}\;,$$ and $\eta$ is the symmetric mass-ratio, $$\eta = \frac{\mathrm{m_1m_2}}{M^2}\;.$$
The amplitudes are scaled relative to 10 $M$ after the peak luminosity in $\psi^{\,\mathrm{NR}}_{22}$ (Sec. \ref{sec:intro:NR_meets_PT}),
\par Note that the \qnm{} frequencies, $\tilde{\omega}_{lmn}$, are complex, and depend on the remnant \bh{'s} parameters: spin magnitude and mass.
\begin{equation}
	\label{eq:OmegaPT}
	\tilde{\omega}_{lmn} \, \equiv \, \omega_{lmn} \, + \, i/\tau_{lmn}
\end{equation}
In \eqn{\ref{eq:OmegaPT}}, $\omega_{lmn}$ is the \qnm{'s} central oscillation frequency, and $\tau_{lmn}$ the mode's decay time.
Each frequency may be conveniently computed using the mapping between $\eta$ and remnant \bh{} parameters given in Eqs. {\ref{eq:mass_model}} and \ref{eq:spin_model}, or Ref. \cite{Healy:2014yta}, along with the phenomenological fitting formulas\footnote{Note that here $\tilde{\omega}_{lmn}$ are in units of $1/M$ while \cite{Berti:2005ys} reports the unitless $M\,\tilde{\omega}_{lmn}$.} for \qnm{} frequencies in Ref. \cite{Berti:2005ys}.

\subsection{From \qnm{s} and templates to \nr{} ringdown analysis}
\label{sec:intro_QNM_To_RD_Analysis}
%
\par Shortly after Vishveshwara's 1970 discovery that perturbed black holes dissipate energy via gravitational \rd{}, the study of perturbed \bh{s} began a proliferation that now enables the creation of \gw{} \rd{} templates \cite{Vishv:1970,Berti:2009kk,Caudill:2011kv}.
In 1971 Teukolsky and Press revealed that \rd{} should be well approximated by a sum of eigenfunctions of Teukolsky's master equation which describes first order departures from the Kerr metric \cite{Teu73_1,PreTeu73_2,LivRevQNM}.
For a \bh{} of mass $M_f$ and dimensionless spin parameter,
\[
j_f=\frac{s_f}{M_f^2}\;,
\]
these eigenfunctions are uniquely determined.
Here $s_f$ is the magnitude of the final \bh{} spin vector.
Press later referred to Teukolsky's set of radial, angular, and temporal eigenfunctions as \qnm{s} \cite{LivRevQNM,Berti:2009kk} [\eqn{\ref{eq:PSI4PT}}].
\qnm{s} are multipoles with the usual polar and azimuthal indices, $\ell$ and $m$.
In addition, in loose analogy with acoustic theory, they are also labeled by an \textit{overtone} number, $n = \{0,1,2...\}$, where, as $n$ increases, so does the typical \qnm{} decay rate \cite{leaver85}.
The $n=0$ \qnm{s} are traditionally referred to as  the \textit{fundamental} modes.
\par Given that astrophysical \bh{s} are expected to be described by only mass and spin, the work of developing \gw{} templates that include \rd{} is largely equivalent to modeling the excitations of Kerr \qnm{s} for different progenitor binaries~\cite{Caudill:2011kv,Berti:2007zu}.
This work has largely focused on the most slowly decaying, fundamental \qnm{s}, which correspond to first order departures from the Kerr metric.
\par However, it has been suggested that second order \qnm{s}, resulting from \textit{nonlinear} self-coupling of their first order counterparts, may also be pertinent \cite{Campanelli:1998jv,Okuzumi:2008ej,Nakano:2007cj,Ioka:2007ak,Pazos:2010xf,Zlochower:2003yh}.
Although these second order \qnm{s} have largely been studied for Schwarzschild \bh{s}, where Regge-Wheeler-Zerilli techniques can be directly applied, formal results for the Kerr case do not appear to exceed \cite{Campanelli:1998jv}, wherein the second order contribution's wave equation is derived within the Newman-Penrose formalism.
\par This result demonstrates that the second order wave equation for Kerr, like its Schwarzschild counterpart, is sourced by a quadratic function of the first order modes.
For this reason it is expected that the second order \qnm{s} for Kerr are characteristically similar to those for Schwarzschild \cite{Ioka:2007ak}.
In particular, one might expect to find within \fig{\ref{fig:full_ringdown}} damped sinusoids whose frequencies and decay rates are sums of those from two first order modes\footnote{This is analogous to the anharmonic oscillator, in which the second order oscillation frequency is twice the first order one \cite{Ioka:2007ak}.}.
\par From these considerations it is clear that \pt{} allows for an extremely rich space of possible \rd{} signals. But given that the fundamental modes are the slowest damped, it is not immediately clear that modes beyond the fundamentals are pertinent to modeling of \nr{} \rd{}.
Indeed, the single and two-mode \rd{}-only templates of Ref.~\cite{Caudill:2011kv} only consider fundamental \qnm{s}.
Similarly, studies that focus on linking \qnm{} excitation with initial binary parameters typically focus only on the fundamental modes \cite{Kamaretsos:2011aa,Kamaretsos:2011um,Kamaretsos:2012bs} and, while work on templates that include both merger and \rd{} has found that overtones are required to blend the two regions, a systematic study of overtone excitement is lacking \cite{Pan:2009wj,Schnittman:2007ij,Taracchini:2013rva,Buonanno:2006ui}.
Moreover, there has been no work published on the detection of nonlinear second order \qnm{s} within \nr{} \bbh{} coalescence.
Here, we inform these areas by describing \qnm{} excitation for a series of initially nonspinning, unequal mass \bbh{} systems.
\par For the recovery of these initial parameters precise agreement between template and signal is needed.
Concurrently, only qualitative agreement is needed for detection purposes \cite{Caudill:2011kv,Berti:2007zu}.
Although a full exploration of detection and parameter estimation is beyond the scope of the current study, we note that the richness of possible signals depends not only on the configuration of the initial binary, but also the orientation of the \bh{'s} final spin vector with respect to the observer's line of sight.
\par As an example, consider again \fig{\ref{fig:full_ringdown}}.
Here we see that if this idealized signal is observed along the remnant \bh{'s} final spin axis, $\theta=0$, then the envelope of its time domain behavior appears to be dominated by a single exponentially decaying function, or equivalently, a single \qnm{}; however, if observed at a significant angle with respect to the final spin axis, here $\theta=\pi/3$, then many \qnm{s} may visibly contribute.
In order to model the complexities of these potential signals, we utilize the intersections between \pt{} and \nr{}.
%
\subsection{Numerical relativity meets perturbation theory}
\label{sec:intro:NR_meets_PT}
%
\par \nr{} waveforms are typically decomposed\footnote{
This decomposition is typically done such that the origin is at the initial binary's center of mass. In general, this is not the location of the remnant \bh{} if there is a nonzero recoil velocity. However, for the systems studied here, the typical distance traveled postmerger, is {sufficiently small} compared to the waveform extraction radius, making this initial center of mass location a good approximation for the position of the remnant \bh{}. Nevertheless, as discussed in Sec. ~\ref{sec:PT_Consistency}
, this does potentially introduce detailed effects that may not be inherent to the \rd{} regime.}
into spin weighted-2 spherical harmonics, $_{-2}Y_{lm}(\theta,\phi)$, such that the Weyl scalar $\psi_4$ is given by
\begin{equation}
	\label{eq:PSI4NR}
	\psi_4(t,\theta,\phi,r) = \; \frac{1}{r} \sum_{l,m} \, \psi_{lm}^{\,\mathrm{NR}}(t) \, [_{-2}Y_{lm}(\theta,\phi)]\;.
\end{equation}
For gravitational radiation, the orthogonality of these harmonics in both $\ell$ and $m$ ensures that this is a true spectral decomposition:
\begin{equation}
\label{eq:PSILM}
	\psi_{lm}^{\,\mathrm{NR}}(t) \equiv \, r \int_{\Omega} \, \psi_4(t,\theta,\phi,r) \, _{-2}\bar{Y}_{lm}(\theta,\phi) \, \mathrm{d}\Omega \, .
\end{equation}
Here $_{-2}\bar{Y}_{lm}(\theta,\phi)$ is the complex conjugate of $_{-2}{Y}_{lm}(\theta,\phi)$, and we will focus on $\psi^{\,\mathrm{NR}}_{lm}$, the spherical harmonic multipoles of the Weyl scalar $\psi_4$. The Weyl scalar $\psi_4$ is related to the observable strain via two time derivatives, $\psi_4 = -\ddot{h}$ \cite{Alcubierre.0707.4654}.
\par During \rd{}, this choice of multipolar decomposition effectively casts the radiation as that corresponding to a perturbed {\it nonspinning} \bh{} \cite{Alcubierre.0707.4654}. However the remnant of a \bbh{} merger is typically a spinning \bh{}.
\par For these cases, the \pt{} of Kerr \bh{s} \cite{Leaver:1986gd} yields
\begin{eqnarray}
	\label{eq:PSI4PT}
	\psi_4(t,\theta,\phi) \approx \; \frac{1}{r} \sum_{l,m,n} \, \psi_{lmn}^{\,\mathrm{PT}}(t) \, [_{-2}S_{lm}(j_f\tilde{\omega}_{lmn},\theta,\phi)]
	\\
	\label{eq:PSIPT}
	\psi_{lmn}^{\,\mathrm{PT}}(t) \equiv \, A_{lmn} \, e^{i\tilde{\omega}_{lmn}t}\,,
\end{eqnarray}
where $\tilde{\omega}_{lmn}$ is the complex \qnm{} frequency, $_{-2}S_{lm}$ are the spin weighted \textit{spheroidal} harmonics, and $A_{lmn}$ are the complex \qnm{} amplitudes or \textit{excitation coefficients} whose magnitude is contingent on where $t$ is chosen to be zero \cite{Berti:2009kk,PhysRevD.73.024013,Zhang:2013ksa}.
\par For example, if $t_*$ is the time relative to the peak luminosity of $\psi^{\,\mathrm{NR}}_{22}$, and one considers \rd{} to include $T_0\,\geq\,10\;(M)$ after $t_*=0$, then $t \equiv t_*-10\;(M)$ \cite{Kamaretsos:2012bs}.
Here we consider $t$ to be in units of the initial binary mass, $M$, which is canonically set to unity.
\par Generally, \eqn{\ref{eq:PSI4PT}} is not an equality as power-law tails, of the form $\psi_{\text{tail}} \sim t^{-k}$, are also expected in the postmerger regime \cite{Price:1972pw,Leaver:1986gd}.
While, in principle, these power-law contributions may be significant near the radiation's peak, a host of numerical studies has shown them to be extremely weak throughout the subsequent \qnm{} regime\footnote{
In contrast to the current study, which evolves the full Einstein equations, studies that are able to resolve late-time power-law decay evolve Teukolsy's equation (e.g. \cite{KrivanLaguna96}), which is motivated by first-order departures from the Kerr space-time \cite{PhysRevLett.29.1114}.
} \cite{Leaver:1986gd,PhysRevD76Berti}.
In particular, while all power-law functions decay slower than exponentials, they also require excitation coefficients much larger than those of \qnm{s} to contribute significantly to the waveform.
Therefore there is a heuristic expectation that the power-law tails eventually dominate the postmerger waveform, but only at very late times \cite{Mitsou:2010jv,Scheel:2003vs,Leaver:1986gd,Harms:2013ib}.
Indeed, recent \nr{} codes that focus on \bbc{} have empirically verified this expectation \cite{Schnittman:2007ij,PhysRevD76Berti,Buonanno:2006ui,Kamaretsos:2012bs}.
Numerical studies that focus specifically on solving Teukolsky's equation do find that power-law tails are physically meaningful, but only at late times, and at amplitudes that are very likely inaccessible to codes that solve Einstein's equations in full \cite{KrivanLaguna96,Harms:2013ib}.
\par While the current study, in part, seeks to describe \rd{} in unprecedented detail, we also find that for the systems considered, power-law decay can be neglected.\footnote{This is readily visible in \fig{\ref{fig:single_fit_on_data}}'s lower panel where, if power-law tails did contribute significantly, they would cause a localized feature near zero frequency. }
%
%
\par For simplicity we have written \eqn{\ref{eq:PSI4PT}} as a sum over the first order \qnm{} indices only.
If written explicitly, the second order QNM terms, being proportional to products of two first order \qnm{s}, would be labeled by six indices, $(l_1,m_1,n_1)(l_2,m_2,n_2)$ \cite{Okuzumi:2008ej,Nakano:2007cj,Ioka:2007ak}.
We have also neglected to explicitly write the \textit{conjugate} or \textit{mirror-mode} terms which arise from Teukolsky's azimuthal equation having two linearly independent solutions that, due to nonzero \bh{} spin, are not the complex conjugates of each other \cite{leaver85,Berti:2005ys}.
\par An additional consequence of nonzero \bh{} spin is that the spheroidal harmonics, while orthogonal in $m$, are {\it not} orthogonal in $l$ for the complex \qnm{} frequencies of \rd{}\footnote{
Specifically, we are concerned with spheroidal harmonics with complex frequency and of spin weight $s=-2$, which correspond to exponentially damped time-domain waveforms.\cite{PhysRevD.73.024013,Stewart:1975vg}
}, making a spectral decomposition of the form of \eqn{\ref{eq:PSILM}} not possible.
However, just as the Kerr metric reduces to the Schwarzschild metric for nonspinning \bh{s}, so do the spheroidal harmonics reduce to the sphericals.
Substituting \eqn{\ref{eq:PSI4PT}} into \eqn{\ref{eq:PSILM}} illustrates this point by revealing that the spherical multipoles of \nr{} are each a sum of many spheroidal \qnm{s} where, in the $j \, \rightarrow \,  0$ limit, only the $l \, = \, l'$ term survives
\begin{eqnarray}
	\label{eq:PSINR_PT_SUM}
	\psi^{\,\mathrm{NR}}_{l'm}(t) \, \approx \, \sum_{n,l} \, A_{lmn} \, \sigma_{l'lmn} \, e^{i\tilde{\omega}_{lmn}t}
	\\
	\label{eq:YSPROD}
	\sigma_{l'lmn} \, \equiv \, \int_{\Omega} \, _{-2}S_{lm}(j_f\tilde{\omega}_{lmn},\theta,\phi) \, _{-2}\bar{Y}_{l'm}(\theta,\phi) \, \mathrm{d}\Omega \,.
\end{eqnarray}
This was first noted in 1973 by Press and Teukolsky \cite{PreTeu73_2} who used standard operator perturbation theory to show that
\begin{align}
	\label{eq:YS_SUM}
	_{-2}S_{lm} \; = \; &_{-2} Y_{lm} \, +  \, j_f \, \tilde{\omega}_{lmn} \sum_{l \neq l'} \, _{-2} Y_{l'm} \, c_{l'lm}& \, \nonumber\\ &+ \; O(j_f \, \tilde{\omega}_{lmn})^2\;.&
\end{align}
Here $c_{l'lm}$ are related to the Clebsch-Gordon coefficients \cite{Berti:2005ys,Teu73_1}.
\par
\begin{figure}[htb]
\includegraphics[width=\factor\textwidth]{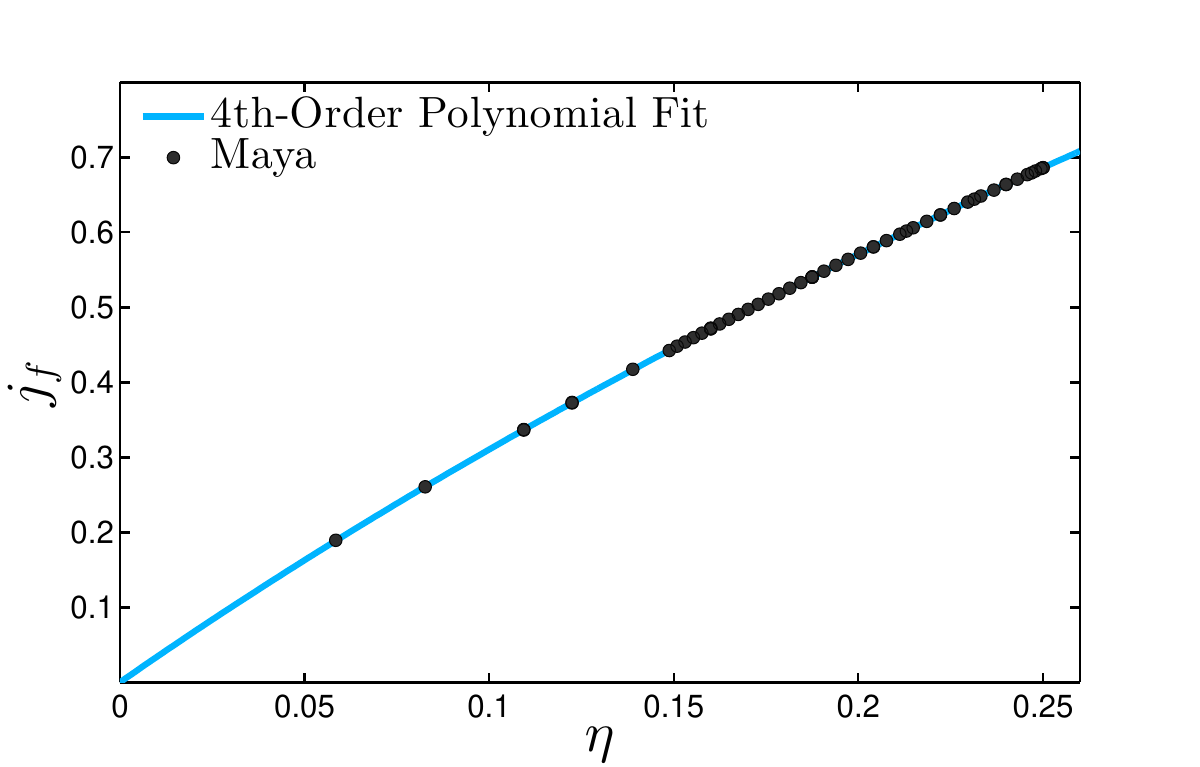}
\caption{Remnant \bh{} spin for initially nonspinning systems of varying mass-ratio. The black dots are final spin values calculated using the isolated horizon formalism \cite{Frauendiener:2011zz}. The trend is monotonic and well fitted with a fourth order polynomial (Appendix \ref{app:mf_jf}).}
\label{fig:jf_on_massratio}
\end{figure}

 Equations~{\ref{eq:PSINR_PT_SUM}} through {\ref{eq:YS_SUM}} motivate two approaches to characterize \qnm{} excitations, $A_{lmn}$: single-mode and multimode fitting.
\paragraph*{\textbf{Single-mode fitting.---}} The first category makes the practical assumption that \eqn{\ref{eq:PSINR_PT_SUM}} is dominated by the $l = l'$ term, and thereby estimates the \qnm{} amplitudes by fitting a single mode to $\psi^{\mathrm{NR}}_{lm}$.
Although this \textit{single-mode} approach has been shown to be effective for the first few $l=m$ multipoles \cite{Kamaretsos:2011aa,PhysRevD76Berti}, in principle, it neglects the presence of overtones and \bh{} spin \cite{Berti:2009kk,Kelly:2012nd}.
Moreover, because \eqn{\ref{eq:YS_SUM}} says that the mixing between spherical and spheroidal harmonics becomes more prevalent for higher spins, we  may hypothesize that single-mode fitting incurs residuals that are qualitatively proportional to the remnant \bh{}'s spin.
In particular, \fig{\ref{fig:jf_on_massratio}} shows that initially nonspinning, quasicircular \bbh{} systems coalesce to form a remnant \bh{s} whose final spin is proportional to the initial binary's symmetric mass-ratio.
We would therefore expect single-mode fitting of these systems to perform better for low mass-ratios ($\mathrm{m_1}\ll\mathrm{m_2}$), and worse at higher mass-ratios ($\mathrm{m_1}\approx\mathrm{m_2}$).
\par Specifically, while it has been shown that \eqn{\ref{eq:YSPROD}}'s $\sigma_{l'lmn}$ can be on the order of 0.10 for moderate values of $j_f$ \cite{PhysRevD.73.024013}, \eqn{\ref{eq:PSINR_PT_SUM}} communicates that the relative values of different $A_{lmn}$ ultimately determine the significance of each \qnm{} term \cite{Kelly:2012nd}.
%
%
\paragraph*{\textbf{Multimode fitting.---}} The second category attempts to fit \textit{each} term in Eqn.~(\ref{eq:PSINR_PT_SUM}), and therefore requires the simultaneous fitting of multiple \qnm{s} within each spherical multipole.
Although this \textit{multimode} approach is more faithful to the fact that the \bh{s} of interest are spinning, current fitting methods have had limited success \cite{Buonanno:2006ui,Berti:2007dg,gr-qc/0608091}.
The difficulty is primarily due to complexity: within each $\psi^{\mathrm{NR}}_{lm}$, a multimode fitting algorithm must optimize over $\lbrace Re[A_{lmn}], \, Im[A_{lmn}], \, \omega_{lmn}, \, \tau_{lmn} \rbrace$ as well as the total number of significant QNMs, $N$.
There are secondary difficulties arising from data accuracy and numerical artifacts.
As a result, the multimode approach is a $4 \times N$ dimensional optimization problem of combinatoric complexity whose solution must be robust against numerical errors.
It is a lot like trying to identify a musical chord by ear.
\phantomsection
\subsection{Structure of the paper}
\label{sec:intro_outline}
\par In the current study we present a multimode fitting method, and apply it to the \nr{} \rd{} of 68 initially nonspinning, unequal mass-ratio binaries with symmetric mass-ratios between $\eta=0.2500$ and $\eta=0.0586$.
\par We report estimates for the \qnm{} excitations of not only fundamental modes, but also for overtones and what appear to be second order modes.
We go on to discuss our results in the context of phenomenological \rd{} models and future detection scenarios.
First, in Sec. \ref{sec:SINGLE_FITS} we review the single-mode approach, and report fit residuals.
As described in Sec. \ref{sec:SINGLE_RES}, for nominal fitting regions, we find that single-mode fitting incurs roughly $1\%$ fitting errors for the best case scenario, and greater $10\%$ error in the worst case scenarios. We also review the systemic dependence of residuals with final \bh{} spin.
In Sec. \ref{sec:MULTI_METHOD} we introduce our multimode fitting method, and compare it with other approaches using mock data in noise, then review found \qnm{} amplitudes and residual errors.
In Sec. \ref{sec:MAPPING}, we present post-Newtonian inspired fits to the dominant \qnm{} excitations across the range of mass-ratios.
In Sec. \ref{sec:PT_Consistency} we discuss the limitations of our results, and their consistency with perturbation theory.
Finally, in Sec. \ref{sec:DISCUSS}, we discuss our results in the contexts of analytic (nonlinear) perturbation theory, and review the significance of our findings to a mock detection scenario.
%
\section{Motivations for multimode Fitting}
\label{sec:SINGLE}
%
\begin{figure*}[htb]
\begin{center}
\begin{tabular}{cc}
\hspace{-1.2cm}\includegraphics[width=\factor\textwidth]{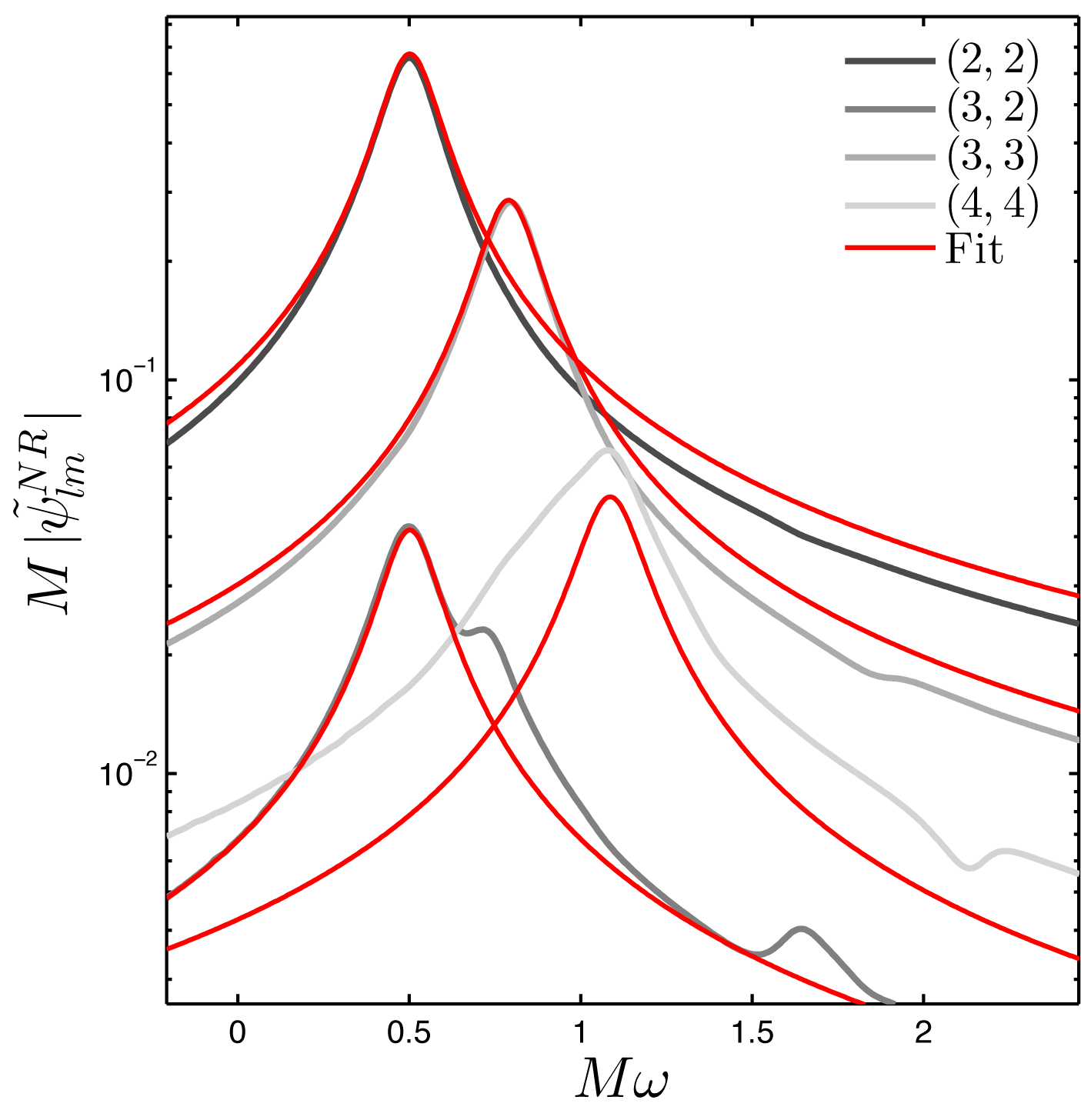}  &
$\;\;\;$\includegraphics[width=\factor\textwidth]{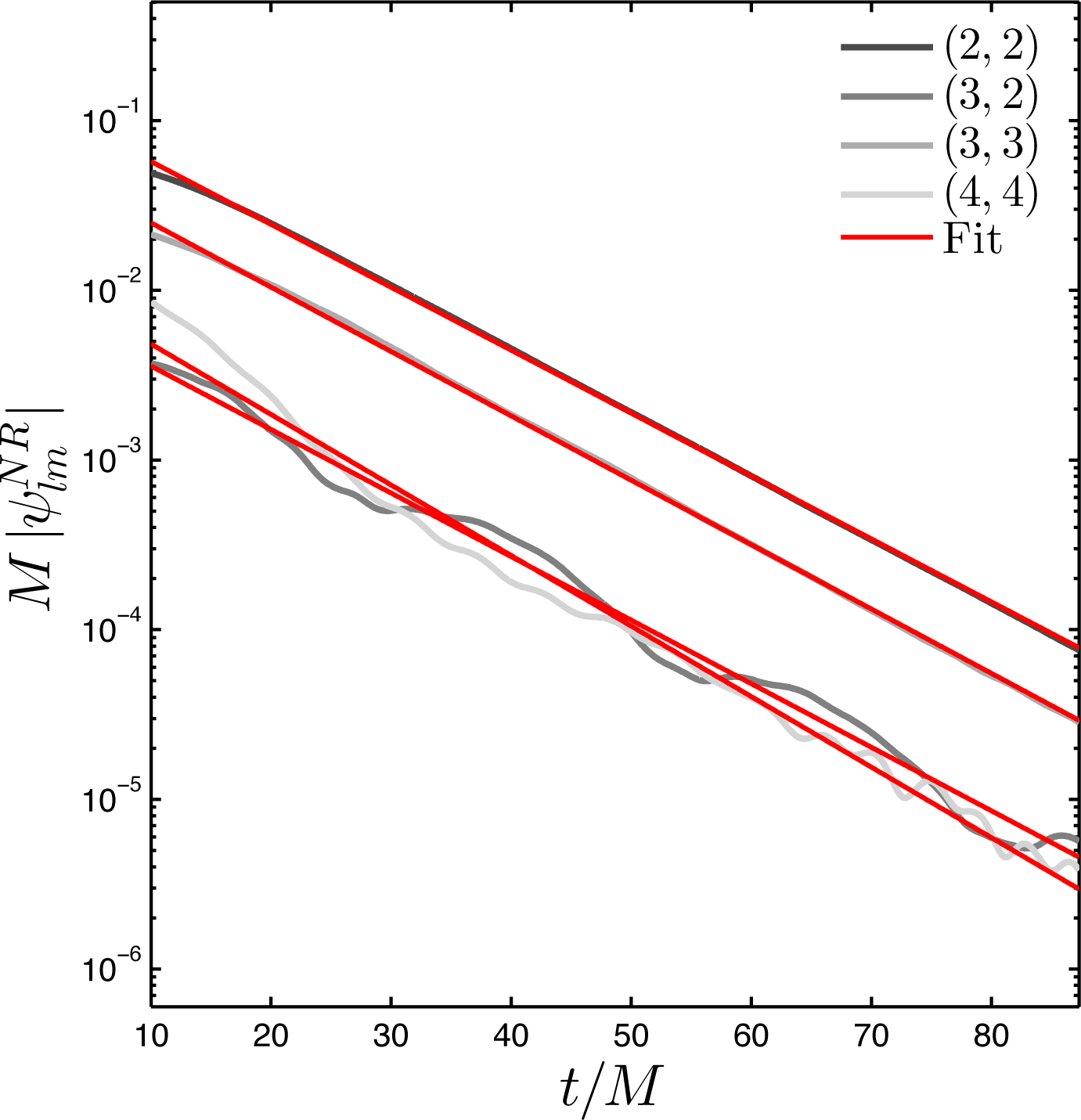} 
\end{tabular}
\caption{ As demonstrated by this set of 2:1 mass-ratio nonspinning waveforms, fitting a single decaying sinusoid to \psinr{l}{m} incurs systematic residuals. \textbf{Top Panel:} The time-domain envelopes for ${(2,2),(3,3),(3,2),(4,4)}$ spherical multipoles and related fits, starting $10M$ after the peak luminosity of \psinr{2}{2}. \textbf{Bottom Panel:} The frequency-domain envelopes, $|\tilde{\psi}_{l,m}^{\mathrm{NR}}|$. All fits correspond to the lowest, $n=0$, \qnm{s}. While the $(2,2)$ and $(3,3)$ multipole waveforms are best described by a single \qnm{} fit, all fits display visible deviations from the raw data.}
\label{fig:single_fit_on_data}
\end{center}
\end{figure*}
\par Let us first consider the single-mode fitting approach discussed in Sec.~\ref{sec:intro:NR_meets_PT}.
Figure~\ref{fig:single_fit_on_data} shows single-mode fits for a 2:1 mass-ratio binary.
While we can see that in this case the subdominant $\psi^{\mathrm{NR}}_{lm}(t)$ are not all simple functions, the dominant multipoles do appear to have exponentially decaying envelopes, and so are well modeled by a single \qnm{}.
Indeed, previous studies have found success in treating the dominant multipoles as single \qnm{s} during \rd{} \cite{Kamaretsos:2011aa,PhysRevD76Berti}.
In particular, this approach has led to effective numerical estimates of \bh{} final spin and mass, as well as the characterizations of fundamental \qnm{} amplitudes with mass-ratio, and initial spin magnitude \cite{Hannam:2010ec,Kamaretsos:2012bs}.
It is therefore fair to suppose that more detailed \qnm{} information is not needed in order to capture \rd{'s} dominant physics.
In what follows, we test this heurism by first outlining the single-mode approach, and then investigating the dependence of fit residuals with initial binary parameters (\fig{\ref{fig:single_fits_and_errors}}).


\subsection{Single-mode fits}
\label{sec:SINGLE_FITS}
\par
First, we outline a qualitatively general single-mode fitting procedure to estimate the fundamental ($n = 0$) QNM excitations:
\begin{enumerate}[a.]
	\item Given the set of $\psi^{\mathrm{NR}}_{lm}$, we define \rd{} to be the region $\{T_0 \leq t \leq T_1 \}$ relative to the peak luminosity\footnote{
	As will be discussed in Sec.~\ref{sec:MULTI_METHOD}, we consider multiple fitting regions in order to characterize both the data and fit.
	In the case of single-mode fitting, fitting regions were chosen to encompass between 86 and 74 $(M)$.
	For the multimode fitting approach to be discussed in Sec.~\ref{sec:MULTI_METHOD}, each waveform was windowed and padded after the onset of numerical noise to maintain a consistent frequency domain resolution.}
	  of $\psi^{\mathrm{NR}}_{22}$ \cite{Kamaretsos:2011aa}.
	\item To calculate the waveform's phase, $\theta_{lm}(t)$, and envelope, $\Psi_{lm}(t)$, we then consider the standard representation for the fit:$$\psi^{Fit}_{lm}\vert_{\{T_0 \leq t \leq T_1 \}} \, = \, \Psi_{lm} \, e^{ i\,\theta_{lm}}\;.$$
	\item We then use linear least-squares fitting to model $\theta_{lm}(t)$ and $Log[\,\Psi_{lm}(t)\,]$ as lines in the time domain:
	\begin{eqnarray}
		\theta_{lm} \, = \, t\,\omega^{Fit}_{lm} + \delta_{lm}^{Fit}
		\\
		Log[\,\Psi_{lm}(t)\,] \, = \, -t / \tau^{Fit}_{lm} \, + \, \text{Log}|A^{Fit}_{lm}|
	\end{eqnarray}
	where $\delta^{Fit}_{lm}$ is the complex phase of $A^{Fit}_{lm}$.
	\item Upon calculating the fit parameters, $\{A^{Fit}_{lm},\omega^{Fit}_{lm},\tau^{Fit}_{lm}\}$, we calculate the \textit{fractional root-mean-square error},
		\begin{equation}
			\label{eq:RMSE}
			\varepsilon_{lm} \, \equiv \,
				\left|   \frac{
					 \langle \; ( \, \psi^{\mathrm{NR}}_{lm} - \psi^{Fit}_{lm} \, ) ^2 \;\rangle }
					{
					 \langle \, {\psi^{\mathrm{NR}}_{lm}}^2 \, \rangle    						} \, \right|   ^ {1/2}.
		\end{equation}
	Here $\varepsilon_{lm}$ is typically much less than $1$ for good fits, and of order $1$ or greater for poor fits.
	More carefully, as discussed in Sec.\ref{sec:SINGLE_ERR}, $\varepsilon_{lm}$ is susceptible to being biased by numerical noise.
	In the worst case scenario, where noise dominates the data to be fit, $\varepsilon_{lm}\,\approx\,1$ may correspond to a minimum residual with respect to fit parameters.
\end{enumerate}
\par Typical single-mode fits are shown in \fig{\ref{fig:single_fit_on_data}} for a $2:1$ mass-ratio binary, with the fitting region starting $T_0 = 10\,M$ after the peak luminosity in $\psi^{\mathrm{NR}}_{22}$.
Here, as well as throughout this paper, the Fourier transform of waveforms, $\psi(t)$, will be denoted as $\tilde{\psi}(\omega)$.
 Note that the $l=m$ multipoles are well fit, with associated errors $\varepsilon_{lm} \approx 0.08$.
However, a notable exception is the $l=m=4$ multipole with $\varepsilon_{44}$ and order of magnitude higher at $\approx 0.65$.
\par Moreover, as has been found in previous studies, we also find that the $l \neq m$ multipoles are generally not well fit by a single \qnm{}.
For example, the $(l,m)=(3,2)$ multipole, $\psi^{\,\mathrm{NR}}_{32}$, is known to have a significant contribution from the $(l,m,n)=(2,2,0)$ term in \eqn{\ref{eq:PSINR_PT_SUM}} \cite{Kelly:2012nd,Buonanno:2006ui,PhysRevD76Berti,0707.0301}.
This may be recognized in the lower panel of \fig{\ref{fig:single_fit_on_data}}, where $\psi^{\,\mathrm{NR}}_{32}$ is seen have its dominant peak not at $\psi^{\mathrm{PT}}_{32}$'s central frequency\footnote{The central frequency is given by the real part of the \qnm{} frequency.} of $M\omega=0.73$, but at $M\omega=0.50$, directly under the peak of $|\tilde{\psi}^{\,\mathrm{NR}}_{22}|$.
\par In what follows we discuss the residual error of the single-mode approach. In particular, we ask if the errors are dominated by numerical artifacts (e.g. resolution related errors \cite{Hannam:2010ec}), or if the errors are dominated by the effects of nonzero \bh{} spin.

\subsection{Single-mode fits: Results and residuals}
\label{sec:SINGLE_RES}
To investigate the residuals incurred by single-mode fitting, we consider 36 initially nonspinning, unequal mass binaries with $\eta$ between $0.2500$ and $0.0586$.
The left panel of \fig{\ref{fig:single_fits_and_errors}} shows typical fit excitation amplitudes, $|A^{Fit}_{lm}|$, and the right panel shows the corresponding residual errors (\eqn{\ref{eq:RMSE}}).
The left panel of \fig{\ref{fig:single_fits_and_errors}} shows that \qnm{} excitation appears regular with symmetric mass-ratio with the $n=0$ mode dominating.
The fitting model proposed in Ref. \cite{Kamaretsos:2011aa} is also plotted.
The lower left panel of \fig{\ref{fig:single_fits_and_errors}} indicates that the $(\ell,m,n)=(4,4,0)$ has a significant local minimum at $\eta\approx0.22$ ($\mathrm{m_1/m_2}\approx 2$) for the resolution in $\eta$ considered here.
The $(\ell,m,n)=(3,2,0)$ \qnm{} has been found to exhibit a similar local minimum \cite{Kelly:2012nd}.
\par Turning to the right panel of \fig{\ref{fig:single_fits_and_errors}},  the $(\ell,m,n)=(2,2,0)$ and $(3,3,0)$ cases show monotonically decreasing trends.
This trend may be due to the difference between spherical and spheroidal harmonics, which is proportional to final \bh{} spin [\eqn{\ref{eq:YS_SUM}}], and is therefore also proportional to symmetric mass-ratio (\fig{\ref{fig:jf_on_massratio}}); thus, single-mode fitting may incur systematic errors that decrease with $\eta$.
\par While the $\varepsilon_{21}$ and $\varepsilon_{44}$ estimates display a more complicated behavior, their overall decrease with $\eta$ suggests that these cases may be significantly affected not only by \qnm{s} beyond the fundamentals, but also by other sources of errors.
%

\subsection{Sources of error}
\label{sec:SINGLE_ERR}
\begin{figure*}[ht]
\begin{center}
\begin{tabular}{cc}
\includegraphics[width=\factor\textwidth]{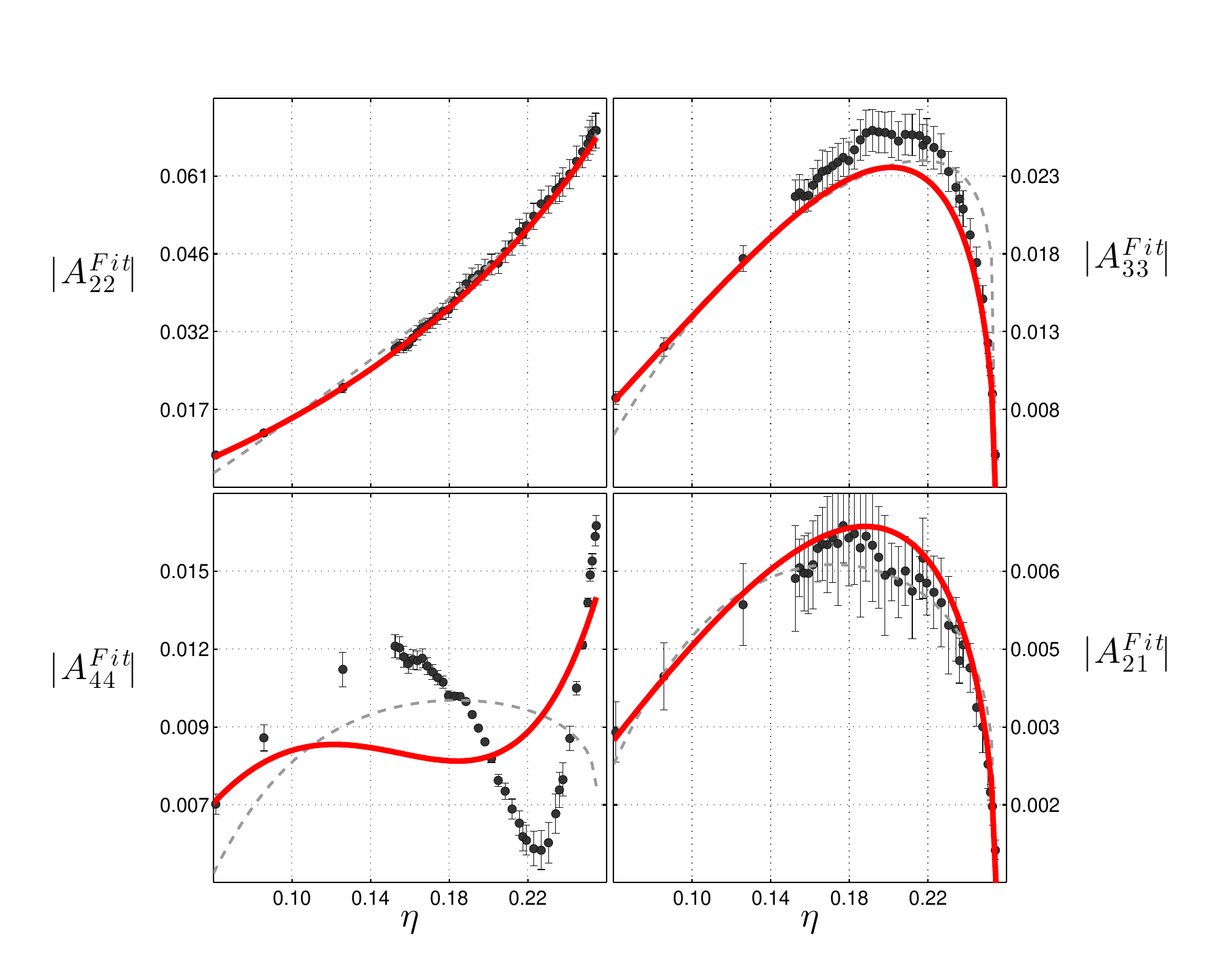} &
\includegraphics[width=\factor\textwidth]{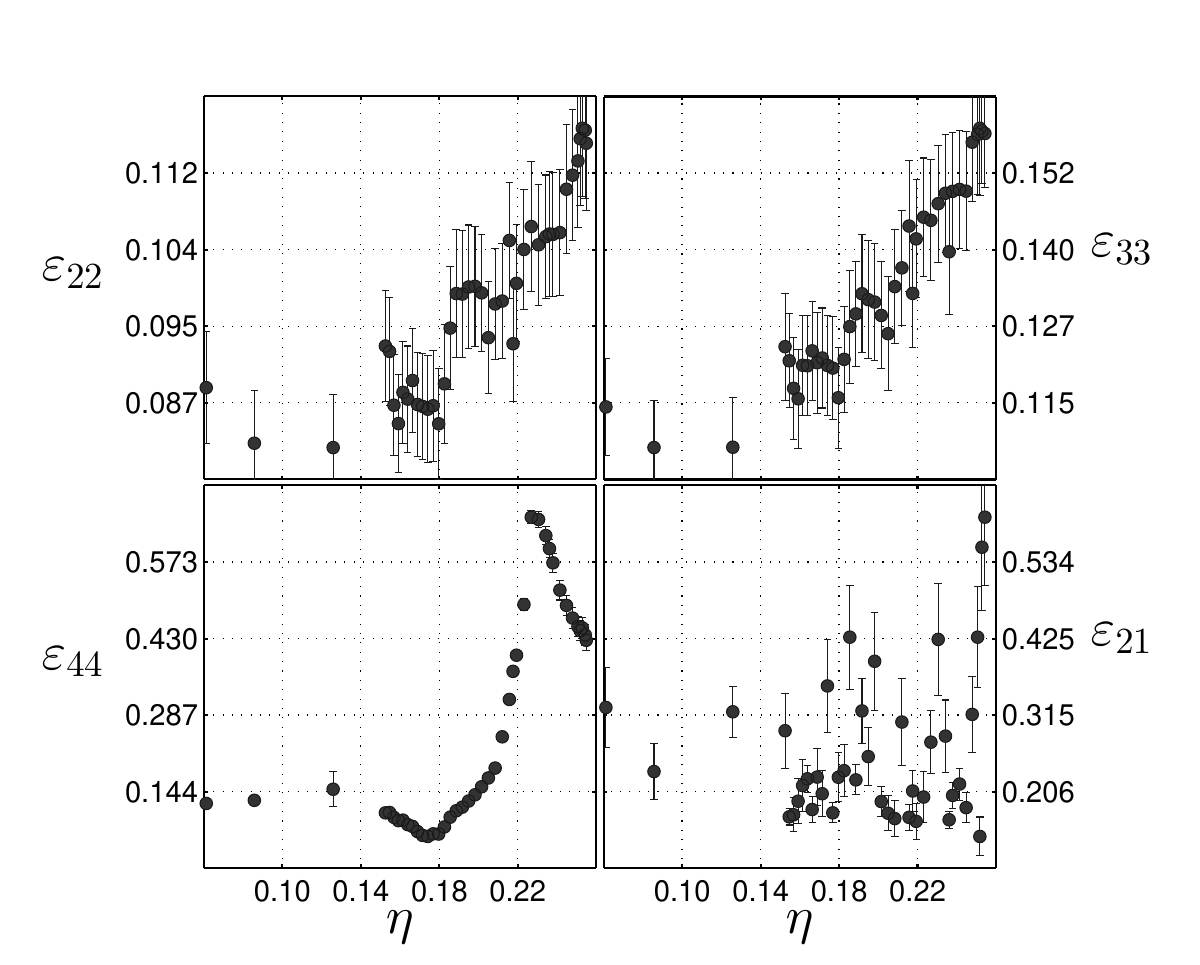}
\end{tabular}\caption{\label{fig:single_fits_and_errors} Here we see the fundamental \qnm{} excitations estimated by single-mode fitting. Left: The black dots are the excitation amplitudes estimated from fitting. For reference, the dashed grey lines are phenomenological fits from Kamaretsos\textit{ et. al. }\cite{Kamaretsos:2011aa}, and the solid red lines are phenomenological fits from the more recent study by Meidam \textit{ et. al.}\cite{Meidam:2014jpa}. The error bars were calculated as described in Sec.~\ref{sec:MULTI_METHOD}- \ref{note:error_bars}. The right set of panels shows the related fractional residual errors calculated via \eqn{\ref{eq:RMSE}}.}
\end{center}
\end{figure*}
To contrast how much of each $\varepsilon_{lm}$ is attributed to nonfundamental \qnm{s} rather than other factors, we briefly review the primary numerical sources of error: finite resolution and extraction radius.
In aggregate, we find that the overall effect of these errors contributes to a noise floor that, at $\sim 10^{-6}\;(1/rM)$, is typically 2 orders of magnitude lower than the relative fit errors shown in the right panel of \fig{\ref{fig:single_fits_and_errors}}.
As a general consequence, $\varepsilon_{lm}$ is increasingly biased by numerical noise as $|\psi^{\,\mathrm{NR}}_{lm}|$ approaches the noise floor.
This is most evident for $\varepsilon_{44}$, which displays a pronounced increase as $|A^{Fit}_{44}|$ sweeps through its local minimum.
\par For the waveforms used here, the simulation grid is structured so that there is a central grid of maximal resolution within peripheral grids whose resolution decreases by a factor of 2 at each outward extension.
The result is an inherent tension between the finite extraction radius, and the finest grid resolution (see Sec. \ref{sec:Result_Limitations} for an expanded discussion of finite extraction radius and related gauge effects.).
In effect, this means that $\psi^{\mathrm{NR}}_{lm}$ up to $\ell=m=5$ are resolved spatially, with $\sim 7$ points oscillation cycle, and temporally with $\sim 42$ points per cycle\footnote{
These figures were calculated using the $\ell=m=5$ \qnm{} frequency for an equal mass nonspinning \bbc{}.
In the same case, we find that there are $\sim 111$ points temporally and $\sim 14$ points spatially within the typical amplitude decay rate.
Because \qnm{} frequency decreases as final spin decreases, these numbers increase as the initial binary becomes more unequal (e.g. \fig{\ref{fig:jf_on_massratio}}).}. 
In particular, we find that duplication of \fig{\ref{fig:single_fits_and_errors}} at $\eta=\{0.25,0.19,0.16\}$ is consistent with resolutions \{0.62,1.125,1.25\} times that of the values quoted above, and, therefore, the right panel of \fig{\ref{fig:single_fits_and_errors}} is not dominated by resolution effects.
\par Our post-merger data contain low amplitude, high frequency oscillations that contribute at most 5\% to our estimates of residual error, $\varepsilon_{lm}$, and appear to be an effect of discretization.
This high frequency contribution is visible in \fig{\ref{fig:single_fit_on_data}} as low amplitude features to the right of each central frequency.
While the high frequency of these oscillations means that their contribution to the mean residual difference is small, the magnitude of these oscillations is also marginal across multipoles, and appears at comparable power at the same positive and negative frequency.
As seen in \fig{\ref{fig:single_fit_on_data}}, this frequency varies from multipole to multipole.
Despite their pervasiveness, these features are too high to be pertinent \qnm{} frequencies [Eq. \ref{eq:YSPROD}], and are likely artifacts due to our simulation's containing nonzero power at frequencies beyond the resolvable limit.
Comparison with public \texttt{NINJA} waveforms \cite{Ajith:2012az} reveals that these features show up inconsistently across NR implementations, which suggests that they are both spurious effects due to discretization, and independent of the dominant physics at play\footnote{
Importantly, as will be discussed in Sec. \ref{sec:MULTI_METHOD}, they are also well localized in the frequency domain, which allows us to effectively filter them out during multimode fitting.}. 
\par As a result, we conclude that the fit errors in \fig{\ref{fig:single_fits_and_errors}} are not dominated by numerical artifacts, but instead primarily due to choice of representation: the spherical representation of \eqn{\ref{eq:PSI4NR}}, versus the spheroidal representation of \eqn{\ref{eq:PSI4PT}}. Kelly \textit{et al} recently came to a similar conclusion by considering only the $(\ell,m)=(3,2)$ spherical multipole.
%
\section{Multimode Fitting: From Spherical to Spheroidal}
\label{sec:MULTI_INTRO}
As discussed in the previous section, the single-mode fitting of spherical multipoles, $\psi^{\,\mathrm{NR}}_{lm}$, results in relatively significant residual errors (greater than 5\%) that are systematic in final \bh{} spin.
This spin-systematic behavior verifies the hypothesis encapsulated by \eqn{\ref{eq:PSINR_PT_SUM}}: \nr{} \rd{} is not a single \qnm{}, but a sum of \qnm{s}.
We are therefore motivated to pursue a multimode fitting approach to describe \qnm{} excitations for different mass-ratios.
In particular, we will seek to extract spheroidal information from the spherical harmonic multipoles of \nr{} waveforms.
\par By noting that the general fitting problem is multilinear in the set of decaying sinusoids given by perturbation theory [\eqn{\ref{eq:PSINR_PT_SUM}}], we present a method based upon ordinary linear least-squares fitting (OLS) to estimate spheroidal QNM amplitudes within each spherical multipole.
We find that this particular choice of fitting routine (e.g. the least-squares approach used here) is not as important as its surrounding algorithm which aims to significantly reduce the problem's complexity.
This is, in part, accomplished by utilizing a standard \textit{greedy} algorithm in addition to OLS fitting.
We refer to our approach as the \GOLS{} method.
\par For reference, we test our method with artificial data within artificial numerical noise to present a brief comparison between our \GOLS{} method and the modified Prony method \cite{Berti:2007dg,Osborne95amodified} in Sec.\ref{sec:MULTI_FITs}.
We then present estimates of the \qnm{} excitations due to initially nonspinning \bh{} binaries of variable mass ratio.
%

\subsection{Multimode fitting method}
\label{sec:MULTI_METHOD}

\par We have developed and implemented the following fitting procedure to estimate \qnm{} amplitudes:
\begin{enumerate}[a.]
	\item Given the set of $\psi^{\mathrm{NR}}_{lm}$, we define \rd{} to be the region $\{T_0 \leq t \leq T_1 \}$ relative to the peak luminosity of $\psi^{\mathrm{NR}}_{22}$ \cite{Kamaretsos:2011um}.
	Because the following procedure involves taking the discrete Fourier transform, each \rd{} waveform is appropriately windowed at the noise floor, and padded to ensure consistent frequency domain resolution.
	\item Following \eqn{\ref{eq:PSINR_PT_SUM}}, we assert that \nr{} \rd{}, $\psi^{\,\mathrm{NR}}_{l'm}$, may be well approximated by sum of \qnm{s}. As our numerical waveforms are of limited accuracy, we consider this sum to be finite:
	\begin{align}
		\label{eq:TRANSF_BASIS}
		\psi^{Fit}_{\mathrm{j}}(t) \, &= \; \sum_{k}^N \, A_{k}^{Fit} \, \sigma_{k\mathrm{j}} \, e^{i\tilde{\omega}_{k}t}& \\
		&\approx \;\; \psi^{\,\mathrm{NR}}_{l'm}&
		\nonumber
	\end{align}
	where
	\begin{equation}
		\label{eq:INDEX_SHORTHAND_J}
		\mathrm{j} \, \longleftrightarrow \, \{l',m\}
	\end{equation}
	and
	\begin{equation}
		\label{eq:INDEX_SHORTHAND_K}
		k \, \longleftrightarrow \, \{l,m,n\}.
	\end{equation}
	While \eqn{\ref{eq:TRANSF_BASIS}}'s $A_{k}^{Fit}$ is the estimate \qnm{} amplitude, for notational simplicity we will henceforth refer to it as $A_{k}$. Moreover, the above summation is only over $\{l,l',n\}$, as $m$ is fixed by \eqn{\ref{eq:PSINR_PT_SUM}}.
	\par Here, the apparent horizon may be used to estimate the \bh{'s} final mass and spin, $M_f$ and $j_f=\frac{s_f}{M_f^2}$ \cite{Frauendiener:2011zz}.
	Alternatively, one may estimate the final \bh{} mass and spin by optimizing the multimode fit of a single $\psi^{\,\mathrm{NR}}_{lm}$, as each \qnm{} frequency is determined by $M_f$ and $j_f$ (Appendix \ref{app:mf_jf}).
	Specifically, the dependence of the \qnm{} frequencies on $M_f$ and $j_f$ may be utilized by either direct calculation (e.g. \cite{leaver85}), as used here, or by phenomenological fit (e.g. \cite{Berti:2005ys})\footnote{We find these two approaches to nominally agree to within 1\% of each other (Appendix \ref{app:mf_jf}). }.
	\item In the language of least-squares fitting, we seek to cast \eqn{\ref{eq:TRANSF_BASIS}} in the form of a set of \textit{normal equations}:
	\begin{eqnarray}
		\label{eq:MULTI_NORMAL}
		\alpha_{\mathrm{i}\mathrm{j}} \, = \, \sum_{k}^N  \, \mu_{\mathrm{i}k} \, \beta_{k\mathrm{j}}
	\end{eqnarray}
	or equivalently,
	\begin{eqnarray}
		\vec{\alpha}_\mathrm{j} \, = \, \hat{\mu} \, \vec{\beta}_\mathrm{j} \; .
	\end{eqnarray}
	To do so, we choose to make the following series of definitions:
	\begin{eqnarray}
		\label{eq:MULTI_BETA}
		\beta_{k\mathrm{j}} \, \equiv \, A_{k} \, \sigma_{k\mathrm{j}}
		\\
		\label{eq:MULTI_ALPHA}
		\alpha_{\mathrm{i}\mathrm{j}} \, \equiv \,
			\frac{1}{\tilde{\omega}_\mathrm{i}} \,
			\int_{T_0}^{T_1} \,
				e^{-i{\omega}_{\mathrm{i}}t} \, \cdot \,
				\psi^{\,\mathrm{NR}}_\mathrm{j}(t) \,
			\mathrm{d}t
		\\
		\label{eq:MULTI_MU}
		\mu_{\mathrm{i}k} \, \equiv
			\frac{1}{\tilde{\omega}_\mathrm{i}} \,
			\int_{T_0}^{T_1} \,
				e^{-i{\omega}_{\mathrm{i}}t} \, \cdot \,
				e^{i \tilde{\omega}_{k} t } \,
			\mathrm{d}t
	\end{eqnarray}
	where $\mathrm{i} \leftrightarrow \{l,m,n\}$ and  $\hat{\mu}$ is an $N \times N$ complex valued matrix.
	The consistency of Eqs. (\ref{eq:MULTI_BETA})-(\ref{eq:MULTI_MU}) with \eqn{\ref{eq:MULTI_NORMAL}} is evident upon plugging \eqn{\ref{eq:TRANSF_BASIS}} into \eqn{\ref{eq:MULTI_ALPHA}}.
	\par If $\hat{\mu}$ is nonsingular, then the complex fitting amplitudes are given by
	\begin{equation}
		\label{eq:MULTI_AFIT}
		\vec{\beta}_{\mathrm{j}} \; = \;\; \hat{\mu}^{-1} \; \vec{\alpha}_\mathrm{j} \; .
	\end{equation}
	\par Recalling that \eqn{\ref{eq:MULTI_BETA}} defines $\vec{\beta}_{\mathrm{j}}$ in terms of the complex \qnm{} amplitudes, we equivalently have that estimates for the spheroidal coefficients in \eqn{\ref{eq:TRANSF_BASIS}} are given by the $k$th element of $\vec{\beta}_{\mathrm{j}}$
	\begin{equation}
		\nonumber
		A_{k} \, \sigma_{k\mathrm{j}} \; = \; (\vec{\beta}_{\mathrm{j}})_k \; = \;\; (\hat{\mu}^{-1}_N \; \vec{\alpha}_\mathrm{j})_k \; .
	\end{equation}
	\par In effect, Eqs. (\ref{eq:MULTI_BETA})-(\ref{eq:MULTI_MU}) entail taking the Fourier transform of the \rd{} waveform, and performing semianalytic, linear least-squares fitting in the basis of damped sinusoids allowed by \pt{}.
	\par This approach imposes that $\psi^{\mathrm{NR}}_{lm}$ be composed of the \qnm{} frequencies of \pt{} rather than treating them as fitting parameters, and therefore, the total dimensionality of the fitting problem is reduced from $4 \times N$ to $2 \times N$: $\{ \, Re[ \beta_{k\mathrm{j}} ], \, Im[ \beta_{k\mathrm{j}}], \, N \}$.
	However, since \eqn{\ref{eq:MULTI_AFIT}} allows for the simultaneous determination of $\beta_{k\mathrm{j}}$'s real and imaginary parts, the problem has effectively been reduced to $1 \times N$ dimensions.
	But note that the problem is not truly linear in $N$, as the fit must be optimized over all likely combinations of \qnm{s} allowed by perturbation theory [\eqn{\ref{eq:PSINR_PT_SUM}}].
	\item To manage this last optimization, we first limit the set of allowed \qnm{s} to those whose $\sigma_{l'lmn}$ is above $5 \cdot 10^{-3}$ [\eqn{\ref{eq:YSPROD}}].
	This choice is practically equivalent to only allowing $l$ to differ from $l'$ by at most 2, and simultaneously limits the largest allowed fitting frequency to be well below that of the non-\qnm{} features discussed in Sec. \ref{sec:SINGLE_ERR}.
	We then use a \textit{greedy}
	\footnote{Our greedy algorithm builds a list of $N$ \qnm{s} by starting with $N = 1$, and adding only \qnm{s} to $\hat{\mu}_N$ that reduce the fit error (\eqn{\ref{eq:RMSE}}). This process continues iteratively until the addition of at most two \qnm{s} does not better the fit significantly, or causes the fit to become worse. A broader description of greedy algorithms may be found in \cite{Cormen2001introduction}.}
	algorithm to estimate the optimal set of $N$ \qnm{s} for each $\psi^{\,\mathrm{NR}}_{lm}$.
	We choose to guide the greedy process by using \eqn{\ref{eq:RMSE}} averaged over different overlapping fitting regions
	\footnote{In particular, we average $\epsilon_{lm}$ over 15 fitting regions whose starting time is equally spaced between $T_0$ and $T_0+20(M )$. Each $\epsilon_{lm}$ is calculated by evaluating \eqn{\ref{eq:MULTI_AFIT}} and \eqn{\ref{eq:RMSE}} on the sub-region.}.
	\item Once the optimal set of \qnm{s} has been found, we estimate the spheroidal \qnm{} amplitudes from Eq. \ref{eq:MULTI_BETA}),
	\begin{equation}
		\label{eq:MULTI_SPHEROIDAL_AMPS}
		 A_{k} \, = \, \frac{ \beta_{k\mathrm{j}} } { \sigma_{k\mathrm{j}} } \; .
	\end{equation}
	\item To quantify the effect\footnote{Please see Sec. \ref{sec:PT_Consistency:FittingRegion} for a somewhat expanded discussion.} of $T_0$ on $A_k$, we perform the above process for $T_0 = \{6,7,8,..11,12\}(M )$ and then rescale each $A_k|_{T_0}$ using the corresponding \qnm{} decay rate such that $A_k$ is relative to $T_0 = 10(M )$.
	The resulting set, $\{A_{k}\}_{T_0}$, describes how much each recovered $A_k$ agrees with our assumption that the choice of fitting regions corresponds to \qnm{} dominated \rd{}. For example, in the ideal case, where the fitting region contains only \qnm{s}, every element $\{A_{k}\}_{T_0}$ would have the same value.
	\label{note:error_bars}
\end{enumerate}
\par Throughout this paper, we describe the fitting region dependence of our results using error bars of width $\frac{1}{2}\mathrm{Range}(\{A_{k}\}_{T_0})$, where $\mathrm{Range}(\{x_k\}) = \mathrm{max}(\{x_k\})-\mathrm{min}(\{x_k\})$.
In \fig{\ref{fig:single_fits_and_errors}}, a scaling factor of $\frac{1}{6}$ is used.
Error bars for nonamplitude quantities have been calculated in a similar fashion.
We choose to represent the error bars according to the range of values because the data of interest are inherently systematic, not random (Appendix \ref{app:rd_start}).
\par Now, for reference, we proceed by touching base with an alternative multimode approach of interest \cite{Berti:2007dg,0264-9381-28-8-085023}, the modified Prony method \cite{Osborne95amodified}.
%
\begin{figure}[htb]
\includegraphics[width=\factor\textwidth]{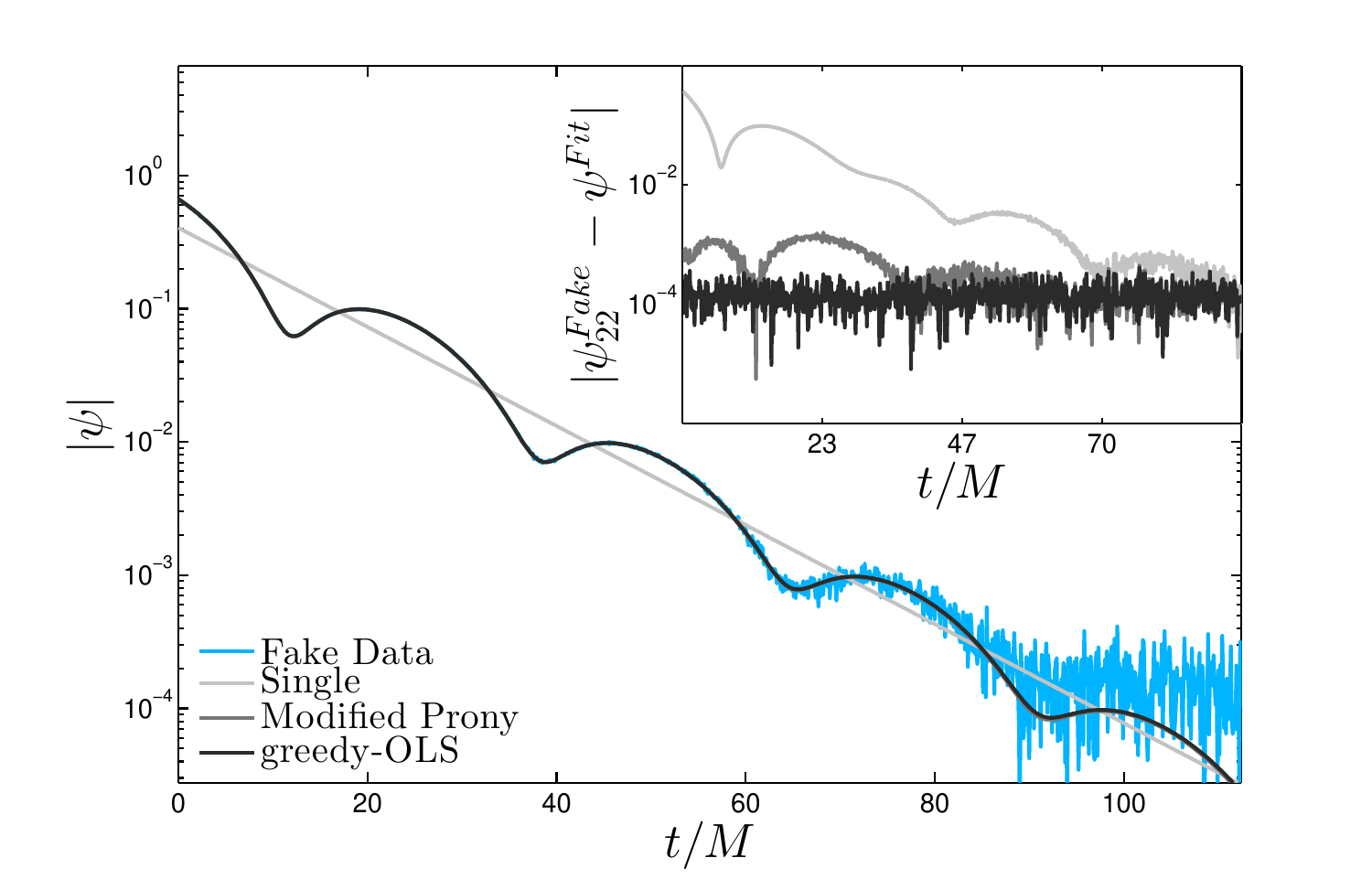}
\caption{ Time domain comparison of different fitting methods for artificial multimode data. }
\label{fig:fit_method_comparison}
\end{figure}
\subsection{Multimode fits}
\label{sec:MULTI_FITs}
%
\par Before using the \GOLS{} algorithm developed in the preceding Sec., we compare it with a popular method for recovering damped sinusoids within noise that linearizes the fitting problem by framing each \qnm{} as the root of a complex polynomial.
If the number of data points is greater than the number of modes, this approach is called the modified Prony algorithm \cite{Osborne95amodified,Berti:2007dg}.
In this Sec. we consider test data to demonstrate what we find to be the typical advantages of approaches like the the \GOLS{} algorithm.
In particular, we ask: given fake data, $\psi^{\,Fake}_{22}$, of known \qnm{} composition, which algorithm returns the input \qnm{s} and achieves the best fit?
\begin{table}[htb]
\caption{\label{tab:fit_comp} Recovered \qnm{s} and errors when applying different fitting methods to artificial \rd{} data composed of the $(\ell,m,n)=\{(2,2,0),(3,2,0),(2,2,1)\}$ \qnm{s} within Gaussian noise. Residual errors were calculated using \eqn{\ref{eq:RMSE}}. }

\centering  
\begin{tabular}{ | l || c | c | }  
  \hline
  \hline
  Method &  Recovered QNMs $(l,m,n)$ & $\varepsilon$  \\[1ex]
  \hline  \hline  Single (Sec.\ref{sec:SINGLE_FITS}) & (2,2,0) & $6.00\times10^{-1}$ \\ 
  \hline
  Modified Prony\cite{Osborne95amodified,Smyth:2003}   & (2,2,0),(3,2,0) & $4.49\times10^{-3}$ \\ 
  \hline
  Greedy-OLS (Sec.\ref{sec:MULTI_METHOD}) & (2,2,0),(3,2,0),(2,2,1) & $1.19\times10^{-3}$ \\ 
  \hline
  \hline
\end{tabular}

\end{table}
\par To portray a typical answer to this question, we construct $\psi^{\,Fake}_{22}$ to be composed of the $(\ell,m,n)=\{(2,2,0),(3,2,0),(2,2,1)\}$ \qnm{s} with the addition of Gaussian noise\cite{Berti:2007dg} that is $10^{-5}$ times smaller than the largest component amplitude.
As the modified Prony algorithm treats \qnm{} frequency and decay time as free parameters, we label each output frequency by its nearest \qnm{} frequency.
\par Figure \ref{fig:fit_method_comparison} compares the output of the \GOLS{} method to the results of the modified Prony algorithm \cite{Osborne95amodified} and the single-mode fitting algorithm described in Sec.\ref{sec:SINGLE_FITS}.
Table \ref{tab:fit_comp} lists the recovered \qnm{s} and corresponding residual errors (\eqn{\ref{eq:RMSE}}).
While both the modified Prony and \GOLS{} methods produce qualitatively precise fits, the inset of \fig{\ref{fig:fit_method_comparison}} shows that the Prony method incurs a noticeably higher residual error.
Turning to Table \ref{tab:fit_comp}, we see that this larger residual error corresponds to the Prony method's not capturing the $(\ell,m,n)=(2,2,1)$ overtone.
This missing mode illuminates two related disadvantages of Prony methods when applied to \qnm{} analysis:  \begin{enumerate}[a.]
	\item The treatment of \qnm{} frequency (\eqn{\ref{eq:OmegaPT}}) as a free parameter increases the difficulty in assigning output frequencies to those predicted by \pt{}.
	\item The method's output frequencies are susceptible to spurious deviations from the structure predicted by \bh{} \pt{}. This aspect of the algorithm complicates the process of estimating \bh{} final mass and spin \cite{Berti:2005ys}.
\end{enumerate}
\par For these reasons, throughout the sections that follow, we favor the \GOLS{} algorithm.
However, we must also note that any fitting algorithm that uses prior information from \pt{} to perform multimode fitting may be just as effective.
For example, we find that using the Levenberg-Marquardt algorithm\cite{marquardt:1963}, in place of \eqn{\ref{eq:MULTI_AFIT}}, is just as potent at estimating the \qnm{} terms in \eqn{\ref{eq:TRANSF_BASIS}}, but only if fitting frequencies are limited to those predicted by \pt{}.
\par
\begin{figure}[htb]
\begin{center}
\begin{tabular}{c}
\includegraphics[width=\factor\textwidth]{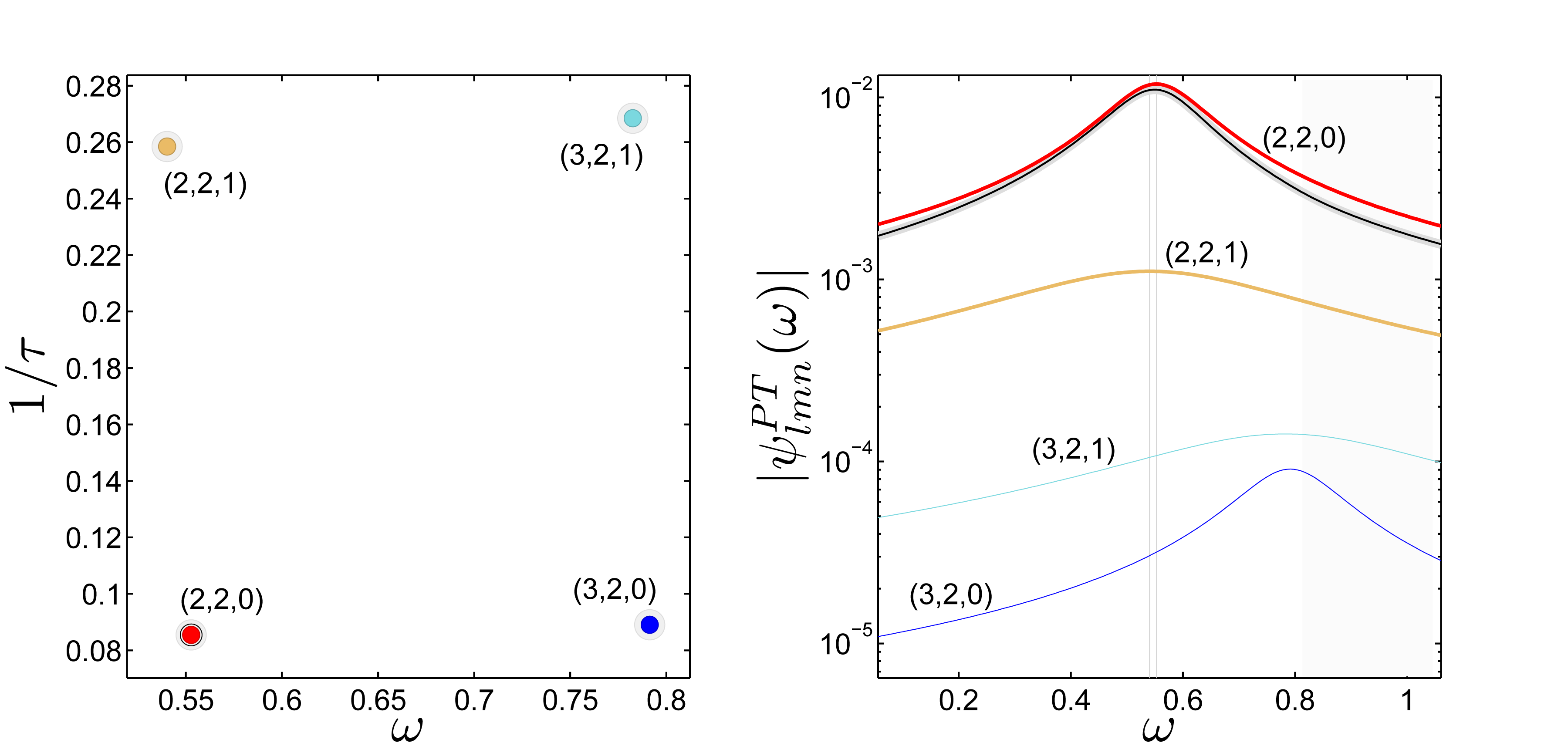} \\
\includegraphics[width=\factor\textwidth]{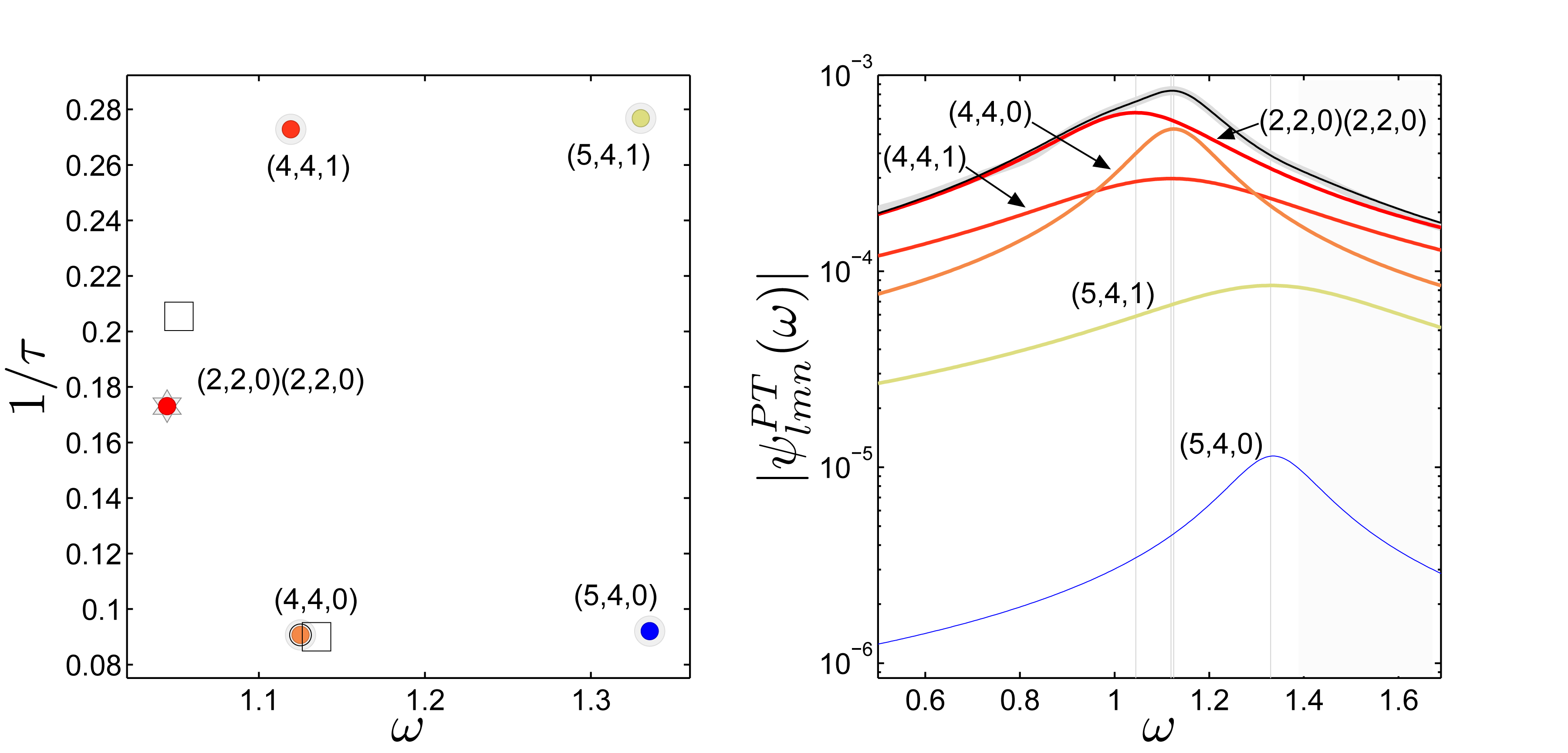}
\end{tabular}
\end{center}
\caption{ Top Panels: multimode fitting results for $\psi^{\,\mathrm{NR}}_{22}$. Bottom Panels: multimode fitting results for $\psi^{\,\mathrm{NR}}_{44}$. Left: \qnm{s} recovered, plotted in central frequency and decay time. Each point is labeled with its \qnm{} index in $(\ell,m,n)$ format. Right: Frequency domain envelopes of component \qnm{s} (color), \nr{} data (grey), and total fit (black). Within each right panel, the shaded region denotes the frequency cut-off. Points in the left panels correspond to curves in the right panels of the same color and \qnm{} label.
For reference, we have overlayed the results of the modified Prony method in \fig{\ref{fig:multi_examples}}'s lower left panel.   }
\label{fig:multi_examples}
\end{figure}

 Now, with some confidence in the \GOLS{} method's faithfulness to the \qnm{} content of \rd{} data, let us consider two applications to \nr{} \rd{}.
Figure \ref{fig:multi_examples} shows results for the $l=m=2$ (top row) and $l=m=4$ (bottom row) spherical multipoles of a 2:1 mass-ratio initially nonspinning \bbh{} system.
%
The four dots in \fig{\ref{fig:multi_examples}}'s top left panel are the recovered \qnm{s} for $\psi^{\,\mathrm{NR}}_{22}$, indicating that $\psi^{\,\mathrm{NR}}_{22}$ is  dominated by four \qnm{s}.
Similarly, $\psi^{\,\mathrm{NR}}_{44}$ appears to be dominated by five \qnm{} terms.
As expected from single-mode fitting, the fundamental modes generally dominate. However, multimode fitting reveals overtones, and in the case of $\psi^{\,\mathrm{NR}}_{44}$, an apparent second order \qnm{}.
For reference, we have overlayed the results of the modified Prony method in \fig{\ref{fig:multi_examples}}'s lower left panel.
\par Importantly, like our test case (Table \ref{tab:fit_comp}), the residual errors for these cases are $\sim10$ times smaller than single-mode fitting. We find this to be generally true for initially nonspinning \bbh{} systems of symmetric mass-ratio between 0.2500 and 0.0586. In the following section, we use these cases to peer into the new information captured by multimode fitting. We model the mapping between initial binary mass-ratio and \qnm{} excitation.
%
%
\section{Mapping QNM Excitation with Symmetric mass-ratio}
\label{sec:MAPPING}

\par We apply the \GOLS{} algorithm to the \rd{} of quasicircular initially nonspinning  \bbh{} systems of symmetric mass-ratio between 0.2500 and 0.0586.
The result is a map between $\eta$ and $A_{lmn}$. Just as in the case of inspiral, with its reflective symmetry about the orbital plane, we find that $|A_{lmn}| = |A_{l-mn}|$ for all systems considered; therefore, we only focus on the $m>0$ multipoles.
\par By applying the \GOLS{} algorithm to our \nr{} \rd{}, we are able to catalog the mass-ratio dependence of overtones and apparent second order \qnm{}.
We find that, for the initially nonspinning systems studied here, the \textit{mirror modes} are not significantly excited.\footnote{We will discuss in Sec.~\ref{sec:PT_Consistency} that imposing these modes detracts from the consistency of our results with \pt{}}
While many well-resolved \qnm{s} are recovered, for practicality, we only focus on those needed to represent $\psi_4$ \rd{} up to marginal accuracy.
We consider these to be \qnm{s} found within the dominant $l=m$ and $l=m+1$ spherical multipoles (e.g. $\psi^{\mathrm{NR}}_{lm}$), where $l \leq 4$ \cite{Kamaretsos:2011aa,0707.0301,Healy:2013jza}.
We go on to present a robust phenomenological model for the mapping between $\eta$ and $A_{lmn}$.
We start by touching base with current models for $A_{lmn}(\eta)$.
\par The phenomenological models proposed by \cite{Kamaretsos:2011aa} are shown in \fig{\ref{fig:single_fits_and_errors}}.
This class of model is derived from the \textit{single-mode} fitting approach mentioned in Sec.\ref{sec:SINGLE}, and only handles $|A_{lmn}|$ while leaving its complex phase to be matched to the phase of $\psi^{\mathrm{NR}}_{lm}$ after merger
\footnote{On the other hand, a multimode representation of each $\psi^{\mathrm{NR}}_{l'm}$ (\eqn{\ref{eq:PSINR_PT_SUM}}) requires information about both $|A_{lmn}|$ and its complex phase}.
While the model functions used in \cite{Kamaretsos:2011aa} capture the qualitative behavior of the first few fundamental QNMs, the current study's increased resolution in mass-ratio reveals clear systematic deviations from \nr{} results (\fig{\ref{fig:single_fits_and_errors}}, left panel).
Most prominently, the local minimum in $|A_{440}|$ is not captured by
\begin{equation}
	\nonumber
	|A_{440}| \, = \, a \; |\tilde{\omega}_{440}|^2 \, \left( \frac{\mathrm{m_1}}{\mathrm{m_2}}\right) ^ {\frac{3}{4}} \, e^{- \, b \; \frac{\mathrm{m_1}}{\mathrm{m_2} }} \, .
\end{equation}
\par The more recent work of \cite{Kelly:2012nd} focuses on the $(l,m,n)=(3,2,0)$ mode, and proposes a qualitatively precise model for $|A_{320}(\eta)|$,
\begin{equation}
	|A_{320}| \, = \, \sqrt{ (a-b\,e^{-\lambda / \eta} )^2 + c^2 } \, ,
\end{equation}
where $a$, $b$, $c$, and $\lambda$ are real valued constants.
Despite the success of this map\footnote{Please see \fig{10} of \cite{Kelly:2012nd}.}, it is not immediately clear why this functional form works so well, and how its effectiveness may be extended to the other \qnm{s}.
\par Ultimately, a thorough analytic study of \qnm{} excitation, akin to \cite{Zhang:2013ksa}, may be needed to  \textit{derive} the mapping between $\eta$ and $A_{lmn}$. While such a pursuit is beyond the current study, a connection between $A_{lmn}(\eta)$ and known physics is appropriate.
\par To approach this problem, we maintain that \qnm{} excitations are, like their \pn{} counterparts, best described by an expansion in the initial binary's parameters.
Here we expand upon \cite{Berti:2007zu} by considering a beyond leading order summation in symmetric mass-ratio.
\begin{figure}[htb]
\begin{center}
\begin{tabular}{c}
\includegraphics[width=\factor\textwidth]{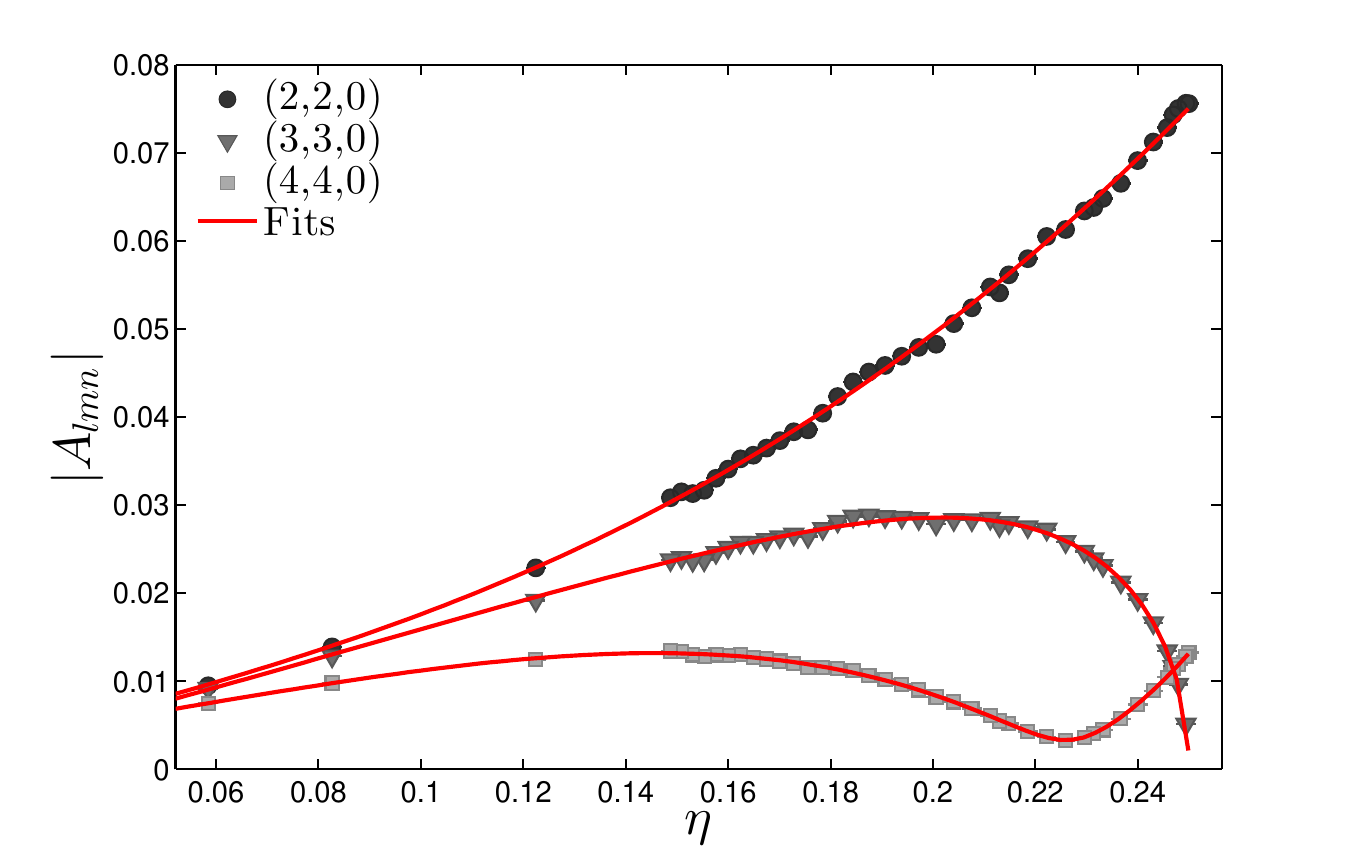}\\
\includegraphics[width=\factor\textwidth]{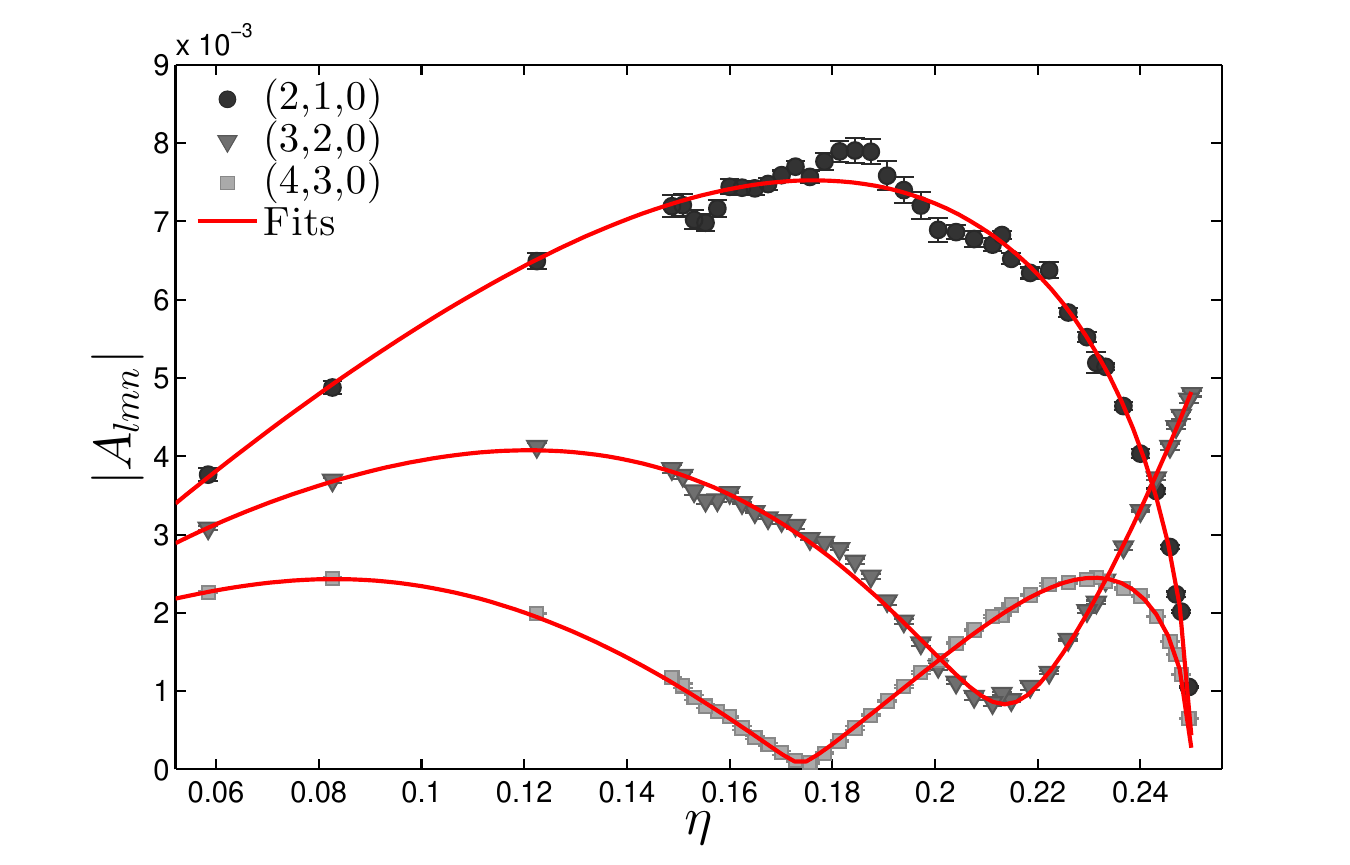}
\end{tabular}
\caption{\label{fig:fundamentals} Fundamentals. The error bars were calculated as described in Sec.~\ref{sec:MULTI_METHOD}- \ref{note:error_bars}.}
\end{center}
\end{figure}
\par First, we note that the relevant\footnote{nonspinning, non-precessing, quasicircular compact binaries.} \pn{} \textit{strain} multipole moments may be written in the form
\begin{align}
\label{eq:PN_Form}
	h_{lm} \, =  \, \eta \, e^{-im\phi(t)} \; \delta_m(\mathrm{m_1},\mathrm{m_2}) \, \sum_{u=0} b_u \, \eta^u \;
\end{align}
where
\begin{align}
	\delta_m(\mathrm{m_1},\mathrm{m_2}) &\, \equiv \,  \frac{|\mathrm{m_1} + (-1)^m\mathrm{m_2}|}{\mathrm{m_1} + \mathrm{m_2}}
	\\ \nonumber
	&\, = \, \sqrt{ 1 - 4\eta  }
\end{align}
and $\phi$ is the time dependent part of the waveform's complex phase \cite{Foffa:2013gja,Kidder:2007rt}.
In seeking to generalize \eqn{\ref{eq:PN_Form}} to $\psi_4$ \qnm{} excitations, we may begin by expecting that during \rd{}, $\phi(t)$ becomes $\phi_{lmn} = \tilde{\omega}_{lmn}t + \, constants$ (we revisit this idea in Sec. \ref{sec:Almn_Fitting}).
Furthermore, since $\psi_4$ and strain are related through two time derivatives, the $\psi_4$ \rd{} analogue of \eqn{\ref{eq:PN_Form}} would pick up a factor of $$\tilde{\omega}_{lmn}^2=|\tilde{\omega}_{lmn}^2|e^{-2\varphi_{lmn}}\;.$$
Lastly, rather than \eqn{\ref{eq:PN_Form}}'s overall scaling by $\eta$, we find it useful to impose that the excitation of each $n$th overtone be proportional to $\eta^n$.
\par Gathering all of these ideas, we propose that, for $\psi_4$ QNM excitations, \eqn{\ref{eq:PN_Form}} generalizes to
\begin{align}
\label{eq:Almn_Model}
	A_{lmn} \, = \;\;& \;  \tilde{\omega}_{lmn}^2 \; \delta_m(\mathrm{m_1},\mathrm{m_2}) \;\eta\, \sum_{u=0} \, a_u \, \eta^u \,&
	\\ \nonumber
	\, = \;\; & e^{-i\phi_{lmn}} \; |A_{lmn}| &
\end{align}
where
\begin{align}
	\label{eq:absolute_phase}
	\phi_{lmn} \; \equiv \; \vartheta_{lmn} \; + \; 2\,\varphi_{lmn}
\end{align}
and
\begin{align}
\label{eq:complex_excitation}
	a_u = |a_u|e^{i\alpha_u}  \; .
\end{align}
\par While we have chosen to encapsulate the intrinsic $\alpha_u$ contribution (\eqn{\ref{eq:complex_excitation}}) within $\vartheta_{lmn}$, one might also expect additional extrinsic contributions to $\vartheta_{lmn}$ from the construction of each simulation (e.g. initial binary separation) \cite{Hannam:2010ec}.
Our approach to these dependencies is outlined in Sec.\ref{sec:Almn_Fitting}.
\par We also notice that our \pn{} inspired model has the immediate advantage of constraining the \qnm{} amplitudes to be zero in the extreme mass-ratio limit, $\eta \rightarrow 0$, while imposing that only even $m$ \qnm{s} are excited in the equal-mass case where $\delta_m = 0$.
As a more phenomenological point, we have chosen to model the overtone dependence as an increasing proportionality in $\eta$ to better fit the \nr{} data.
\par With these conceptual tools at hand, we may now apply \eqn{\ref{eq:Almn_Model}} to \nr{} \rd{} by constructing a fit for the complex valued $A_{lmn}$, as a function of $\eta$.
\subsection{Constructing a fit for $A_{lmn}$ on $\eta$}
\label{sec:Almn_Fitting}
\par In order to accurately model \rd{} according to \eqn{\ref{eq:PSI4PT}}, both $|A_{lmn}|$ and the overall phase, $\phi_{lmn}$ must be represented.
To do so, let us start by focusing on the aspects of \eqn{\ref{eq:Almn_Model}} not given by \pt{}.
First, we rearrange \eqn{\ref{eq:Almn_Model}} to define
\begin{align}
	\label{eq:Almn_Transform}
	C_{lmn} \; \equiv \; &\frac{A_{lmn}}{\eta^{1+n} \;\; \tilde{\omega}_{lmn}^2 \; \delta_m(\mathrm{m_1},
	\mathrm{m_2})}&
	\\ \nonumber
	 = \; &\sum_{u=0} \, a_u \, \eta^u \; .&
	 \\ \nonumber
	 = \; & |C_{lmn}| \; e^{i \, \vartheta_{lmn}} &
\end{align}
As we expect $C_{lmn}$ to be a polynomial with complex coefficients, it might be well captured by standard least-squares fitting methods; however, we are wary that this approach will be ineffective if $\vartheta_{lmn}$ is not dominated by the phase of the polynomial sum\footnote{For simplicity, we will not separate the Kerr eigenvalues (e.g. the excitation factors \cite{Berti:2006wq}) out from the net \qnm{} excitation, $A_{lmn}$. The result is that the polynomial in question approximates the product of two functions. One, the excitation factor, is independent on the initial parameters. The other is entirely dependent on the initial parameters. }.
\begin{figure}[htb]
\includegraphics[width=\factor\textwidth]{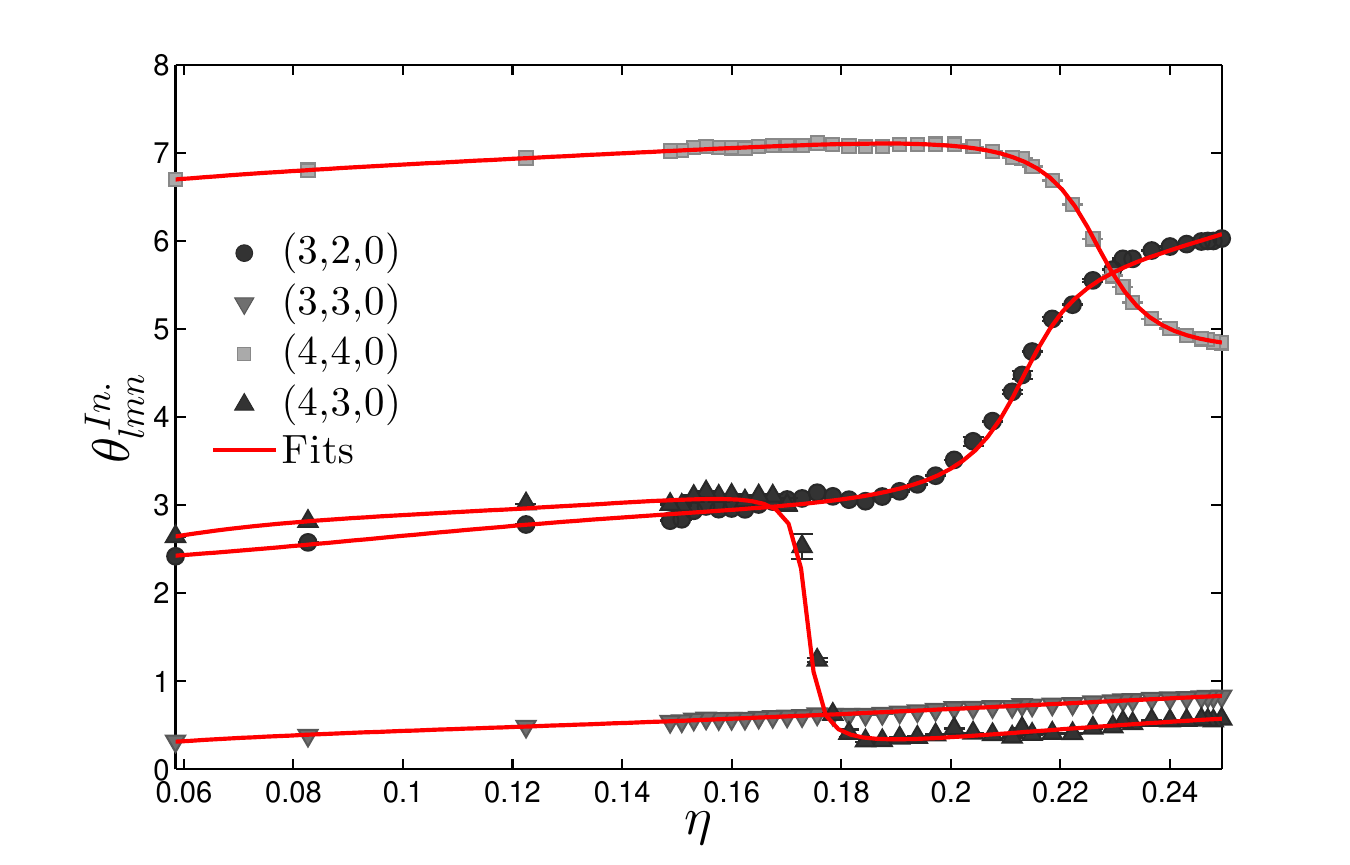}
\caption{\label{fig:relative_phase} Examples of phases relative to $m\phi_{22}/2$.}
\end{figure}
\par With this in mind, if we refer to the intrinsic polynomial phase as $\vartheta^{\,In.}_{lmn}$, and the additional extrinsic contribution as $\vartheta^{\,Ex.}_{lmn}$, then
\begin{align}
	\label{eq:phase_extrensic_and_intrinsic}
	\vartheta_{lmn} \; = \; \vartheta^{\,Ex.}_{lmn} \; + \; \vartheta^{\,In.}_{lmn}\;.
\end{align}
Physically, if there is a preferred azimuthal direction postmerger, then one might expect it to dominate $\vartheta^{\,Ex.}_{lmn}$.
\par In practice, we find this preferred direction is set by the kick velocity.
For the simulations considered here, the kick velocity is always within the orbital plane of the initial binary, giving $\vec{v}_{kick} = v_x \hat{x} + v_y \hat{y}$.
The direction of the kick velocity with respect to the simulation frame is then $\phi_{kick} = \tan^{-1}( {v_y}/{v_x})$.
In this sense, we find that the extrinsic part of $C_{lmn}$'s complex phase is given by
\begin{align}
	\vartheta^{\,Ex.}_{lmn} \; = \; m \, ( \phi_{kick}+\phi_0 ) \; .
\end{align}
Together with \eqn{\ref{eq:phase_extrensic_and_intrinsic}} and \eqn{\ref{eq:absolute_phase}}, we now have that
\begin{align}
	\label{eq:absolute_phase_2}
	\phi_{lmn} \; = \; \vartheta^{\,In.}_{lmn} \; + \; m \, ( \phi_{kick}+\phi_0 ) \; + \; 2\,\varphi_{lmn}  \;.
\end{align}
Note that changes in the line of sight about the \bh{'s} final spin direction affect $\phi_{lmn}$ and $ m \, \phi_{kick}$ in the same way.
Put differently, redefining \eqn{\ref{eq:PSILM}}'s to be $\phi = \phi' - \delta\phi$ effectively adds $m\delta\phi$ to both sides of \eqn{\ref{eq:absolute_phase_2}}.
This leaves \eqn{\ref{eq:absolute_phase_2}}'s $\phi_0$ as an orientation independent quantity (e.g. independent of the observer's location in the initial binaries orbital plane).
\par However, $\phi_0$ is not purely intrinsic.
As we have written it in \eqn{\ref{eq:absolute_phase_2}}, $\phi_0$ not only encapsulates the difference between the final kick orientation and \qnm{} phase, but also how each \qnm{}'s phase has evolved up to the start of the fitting region
, $t_*=T_0$.
This is discussed further in Sec. \ref{sec:Result_Limitations:Start_Time}.
\par Using the $(l,m,n)=(2,2,0)$ \qnm{}, we find that
\begin{align}
\label{eq:phi0_definition}
	\phi_0 \equiv  \;  \frac{ \vartheta^{\,Ex.}_{lmn} }{m} - \phi_{kick}
	\approx \; \frac{ \phi_{220} }{2} - \phi_{kick}  \;.
\end{align}
This gives $\phi_0=-2.39\pm0.10\;rad$.
The regularity of approximation across different mass-ratios is briefly discussed in Sec.~\ref{sec:Result_Limitations}.
\par Together, Eqs.~(\ref{eq:absolute_phase})-(\ref{eq:phi0_definition}) reveal the intrinsic polynomial phase to be
\begin{align}
	\label{eq:intrinsic_phase}
	\vartheta^{In.}_{lmn} \; \approx & \; \phi_{lmn} \; - \; (2\varphi_{lmn}+m(\phi_{kick}+\phi_0)) &
	\\ \nonumber
	\; \approx & \; \phi_{lmn} - \; (2 \varphi_{lmn} + m\frac{\phi_{220}}{2}) \; . &
\end{align}
\par We may therefore construct $C_{lmn}$ by evaluating \eqn{\ref{eq:intrinsic_phase}}, and applying it to the magnitude of $|C_{lmn}|$ given by \eqn{\ref{eq:Almn_Transform}}.
This allows for the \textit{simultaneous} least-squares fitting of $C_{lmn}$'s magnitude and phase. Here we have used \texttt{MATLAB}'s $\mathrm{polyfit.m}$.
By increasing the order of the polynomial fit until the residual error [\eqn{\ref{eq:RMSE}}] changes by less than $10\%$, we find that $C_{lmn}$ are well fit by polynomials of order $\ell-1$ for the considered range of $\eta$.
Figure~\ref{fig:fundamentals} displays the broad effectiveness of our fitting $C_{lmn}$, and then transforming back to $A_{lmn}$ to calculate $|A_{lmn}|$.
Similarly, \fig{\ref{fig:relative_phase}} displays the corresponding intrinsic phases and their fits.
\par For each local minimum in \fig{\ref{fig:fundamentals}}, there is a corresponding \textit{phase transition} in \fig{\ref{fig:relative_phase}}.
In an approximate sense, this suggests that each $C_{lmn}$ may be more appropriately represented as a polynomial function of $(\eta - \eta_0)$, which would force $\eta=\eta_0$ to be a local minimum. However, for simplicity, we have tabulated all fitting coefficients according to \eqn{\ref{eq:Almn_Transform}}.
\par All fitting coefficients are given in Appendix \ref{app:Fit_Coeffs}.
%
\subsection{Beyond the fundamentals: overtones \& second order modes}
\label{sec:Beyond}
%
\par
\begin{figure}[htb]
\begin{center}
\begin{tabular}{c}
\includegraphics[width=\factor\textwidth]{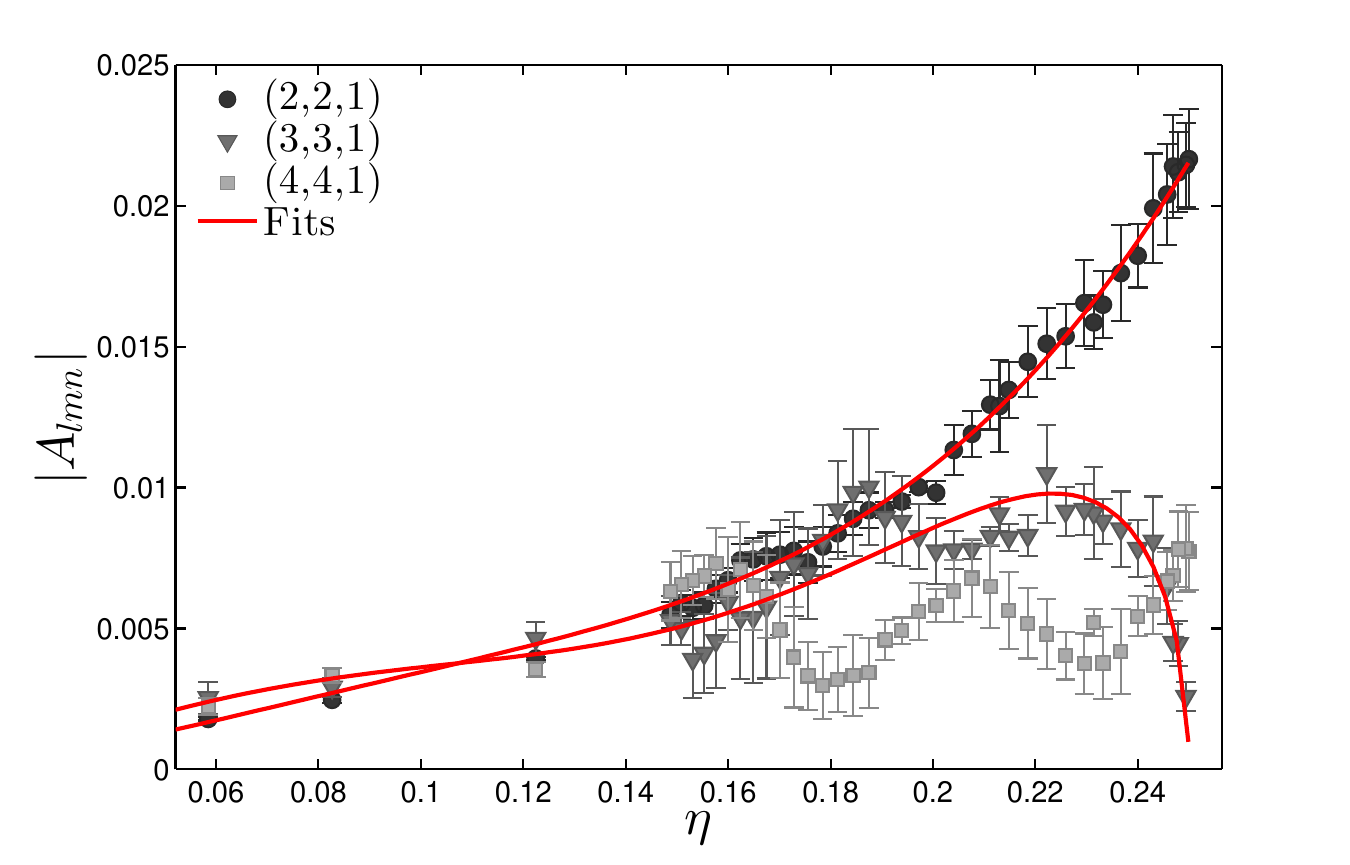}\\
\includegraphics[width=\factor\textwidth]{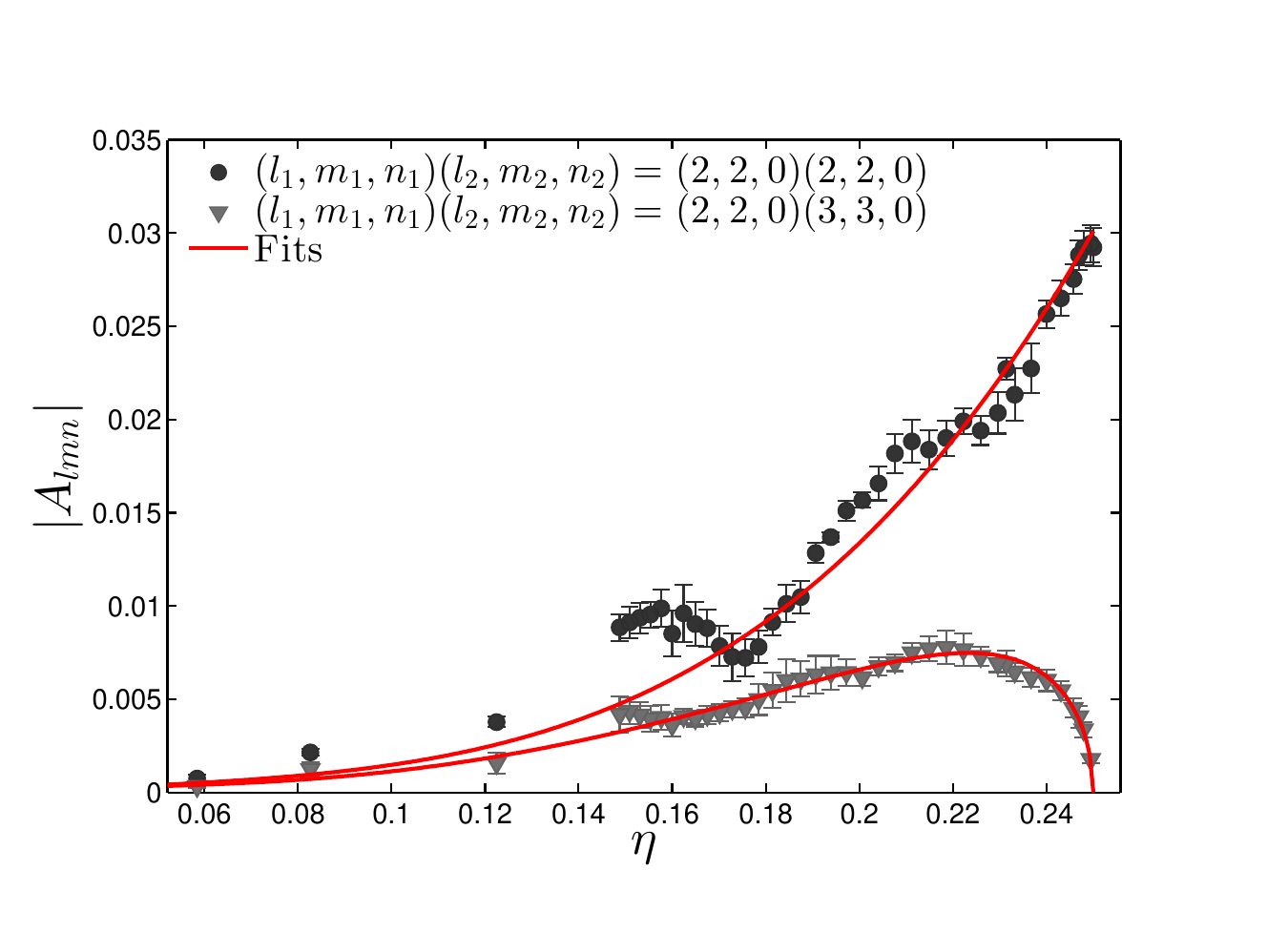}
\end{tabular}
\caption{\label{fig:overtones_and_2ndOrder} Estimated overtone (Top) and second order (Bottom) excitation amplitudes via multimode fitting. The error bars were calculated as described in Sec.~\ref{sec:MULTI_METHOD}- \ref{note:error_bars}.}
\end{center}
\end{figure}

 Figure~\ref{fig:overtones_and_2ndOrder} displays estimates for the \qnm{} amplitudes of overtones (top panel) and second order modes (bottom panel) as recovered by the \GOLS{} algorithm.
While their existence has been discussed in previous studies (e.g \cite{Okuzumi:2008ej,Nakano:2007cj,Ioka:2007ak,Pazos:2010xf,Zlochower:2003yh,Pan:2009wj,Schnittman:2007ij,Taracchini:2013rva,Buonanno:2006ui}), we present for the first time their characterization with symmetric mass ratio.
\par The fitting polynomials for the overtones were found to be of order $l-1$ in $\eta$.
The $(l,m,n)=(4,4,1)$ case is a clear exception, requiring at least an eighth order fit.
While we find that many of our estimates of $|A_{lmn}|$ display a localized increase between $0.18 \geq \eta \geq 0.17$, $|A_{441}|$ displays a significant decrease which makes its $\eta$ dependence possibly inconsistent with \eqn{\ref{eq:Almn_Model}}. As discussed in Sec.\ref{sec:Result_Limitations:Start_Time}, this is likely due to the definition of \rd{} start time in terms of the initial rather than final mass scale.
\par Given the limitations of our \nr{} runs, we consider these oscillations to be numerical, rather than physical.
A similar oscillating trend is observed in the apparent $(l_1,m_1,n_1)(l_2,m_2,n_2) \, = \, (2,2,0)(2,2,0)$ excitation (\fig{\ref{fig:overtones_and_2ndOrder}}).
We discuss the likely source for these oscillations in the next section (Sec. \ref{sec:Result_Limitations}).
\par While the overtones decay faster (e.g. \fig{\ref{fig:multi_examples}}), their functional form largely mirrors their $n=0$ counterparts (\fig{\ref{fig:fundamentals}}).
Similarly, the functional form of the second order modes appears consistent with the notion that each second order mode is largely driven by products of two first order modes \cite{Ioka:2007ak}. Quantitatively, we expect that each $A_{(l_1 m_1 n_1)(l_2 m_2 n_2 )}$ should be proportional to the product of some $A_{l_1 m_1 n_1 }$ and  $A_{l_2 m_2 n_2 }$
\begin{align}
	A_{(l_1 m_1 n_1 )(l_2 m_2 n_2 )} \; \propto \; &\,A_{l_1 m_1 n_1 }\;A_{l_2 m_2 n_2 }\,& \; .
\end{align}
Under this caveat, we model the second order modes according to
\begin{align}
\label{eq:second_order_Alm_model}
	A_{(l_1 m_1 n_1 )(l_2 m_2 n_2 )} \; = \; &\mu_{(l_1 m_1 n_1 )(l_2 m_2 n_2 )}&
	\\ \nonumber
	&\times \,A_{l_1 m_1 n_1 }\,\;\,A_{l_2 m_2 n_2 }\, \;, &
\end{align}
where, given $A_{l_1 m_1 n_1 }$ and $A_{l_2 m_2 n_2 }$ from the first order fits, $\mu_{(l_1 m_1 n_1 )(l_2 m_2 n_2 )}$ is the only undetermined parameter.
\par Upon using a standard root finding algorithm to solve for $\mu_{(l_1 m_1 n_1 )(l_2 m_2 n_2 )}$, we find qualitatively good agreement between our raw estimates for $A_{(l_1 m_1 n_1)(l_2 m_2 n_2 )}$ and \eqn{\ref{eq:second_order_Alm_model}}.
While \fig{\ref{fig:overtones_and_2ndOrder}} displays $(l_1 m_1 n_1)(l_2 m_2 n_2 )  =  (2,2,0)(2,2,0)$ and $(2,2,0)(3,3,0)$ cases, other less dominant and poorly resolved candidates were detected.
\par All fitting coefficients are given in Appendix \ref{app:Fit_Coeffs}.
%
\section{Consistency with Perturbation Theory and Result Limitations }
\label{sec:PT_Consistency}
%
\begin{figure*}[t]
\begin{center}
\begin{tabular}{cc}
\multicolumn{2}{c}{ \includegraphics[width=\factor\textwidth]{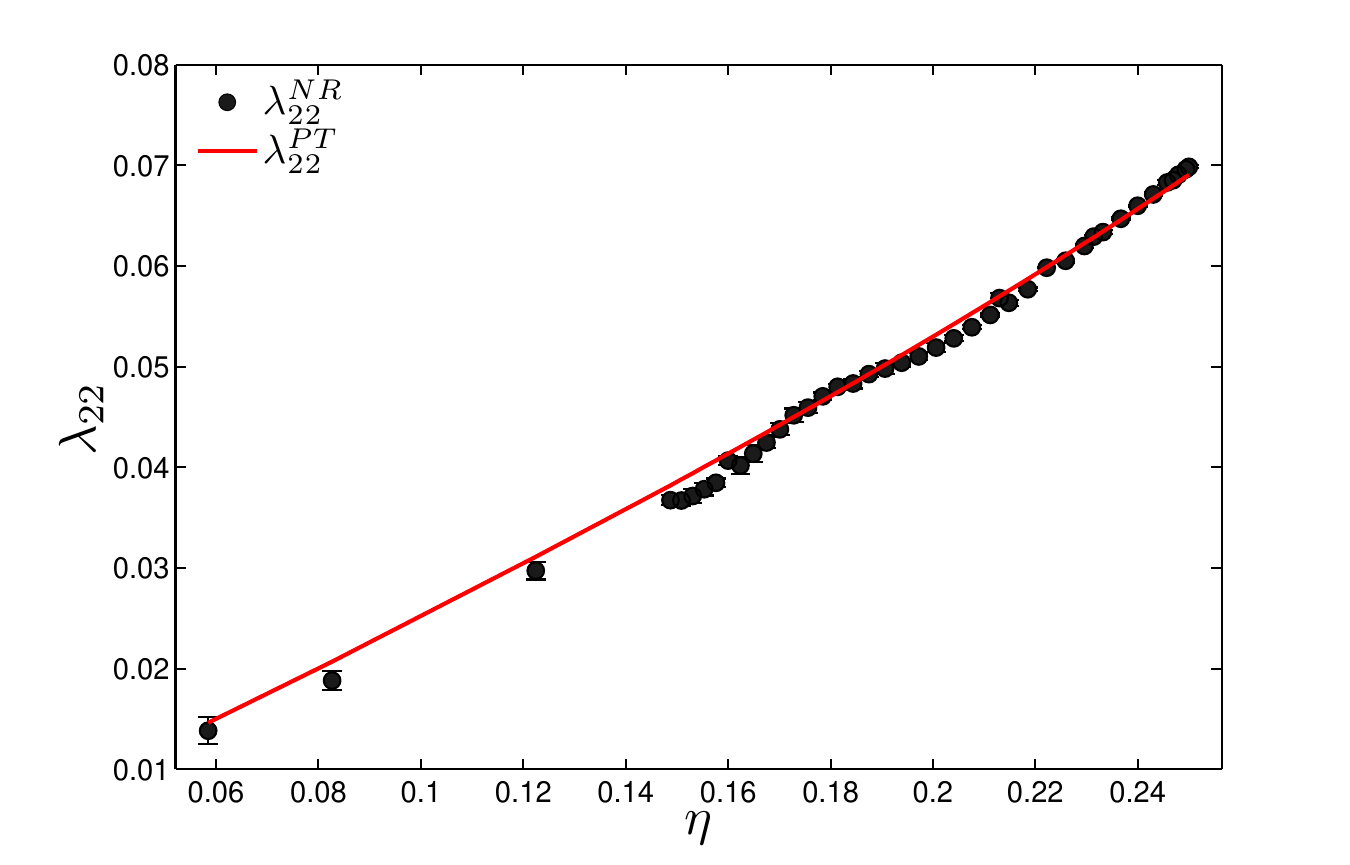} }
\\
\includegraphics[width=\factor\textwidth]{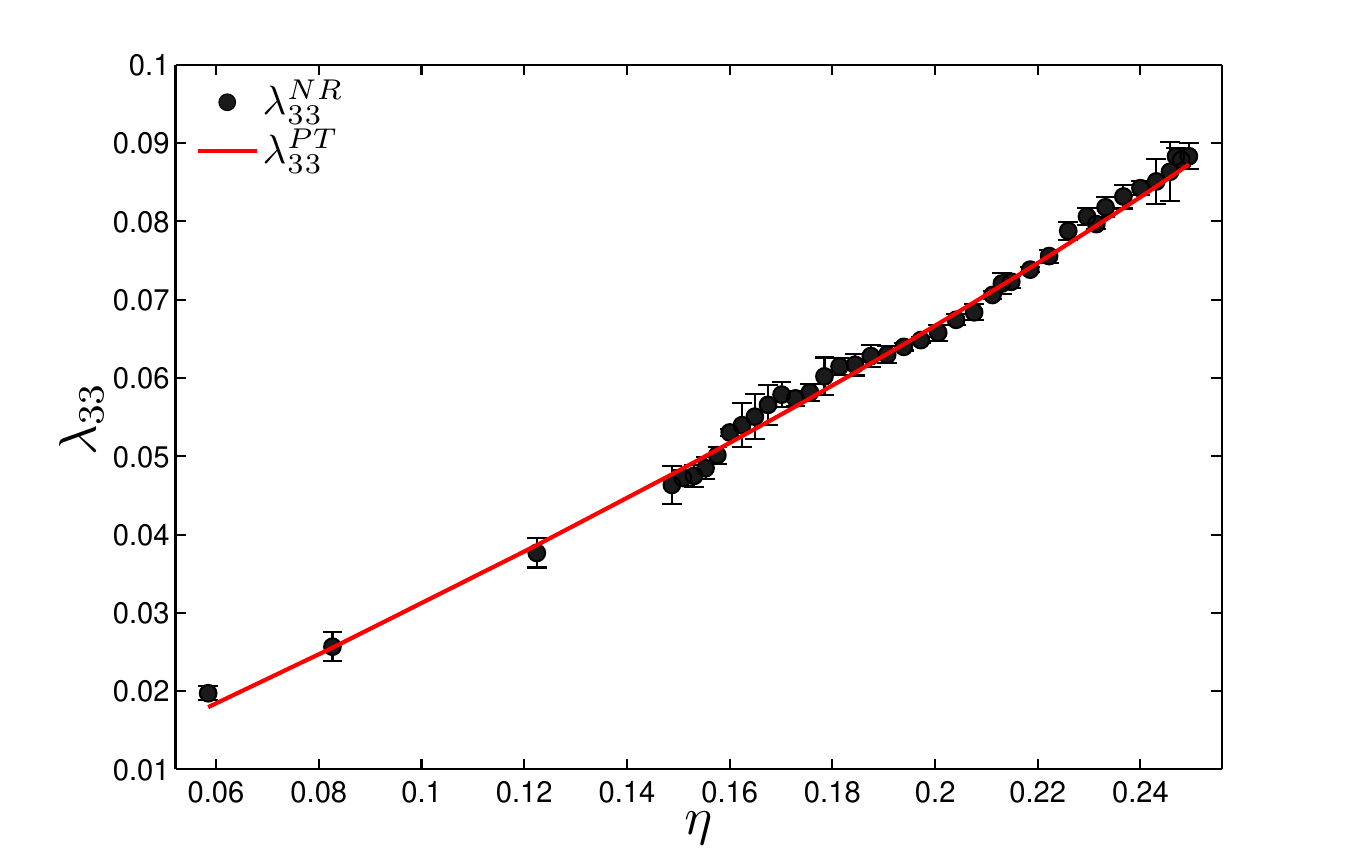} &
\includegraphics[width=\factor\textwidth]{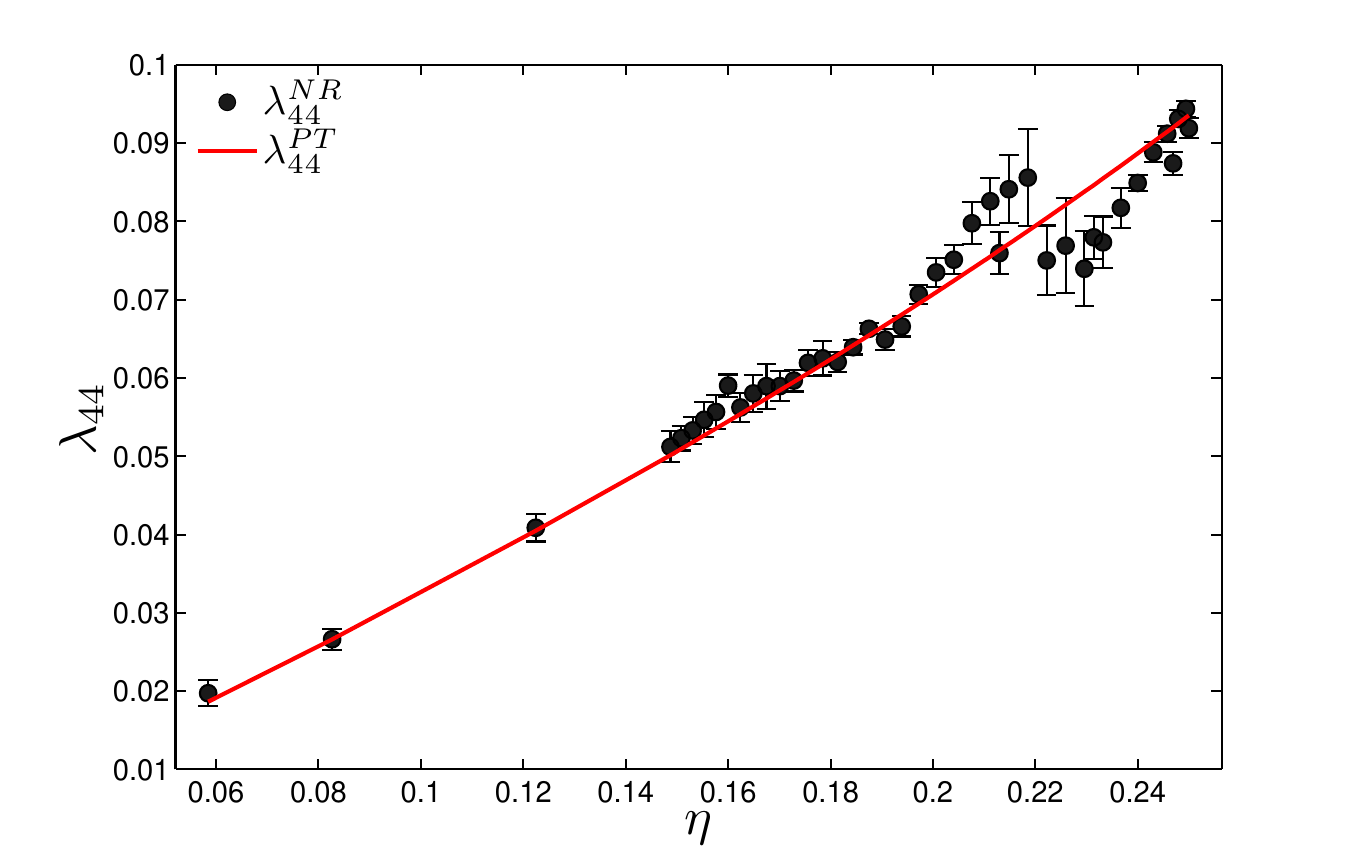}
\end{tabular}
\caption{\label{fig:product_ratios} {Top, bottom left, right:} Ratio of inner-products between spherical and spheroidal harmonics estimated via multimode fitting and direct calculation. The error bars were calculated as described in Sec.~\ref{sec:MULTI_METHOD}- \ref{note:error_bars}.}
\end{center}
\end{figure*}
\par While we have developed a method for the estimation of \qnm{} excitation coefficients, this alone does not guarantee the consistency of our results with perturbation theory.
This is primarily due to the fact that the \qnm{s} and their related functions are not complete (e.g. \cite{gr-qc/9810074}).
In particular, the decaying sinusoids are overcomplete, making it, in principle, possible to achieve an arbitrarily good fit to \eqn{\ref{eq:PSINR_PT_SUM}} with many different combinations of decaying sinusoids.
However, the effectiveness of the \GOLS{} method described in Sec.~\ref{sec:MULTI_INTRO} hinges not on the completeness of the \qnm{s}, but on the uniqueness of the Fourier transform (\eqn{\ref{eq:MULTI_ALPHA}}), which the algorithm seeks to approximate up to numerical accuracy by focusing only on the sparse \qnm{} frequencies suggested by \pt{}\footnote{
The \GOLS{} algorithm uses only a handful of frequencies to estimate the Fourier Transform at \textit{all} frequencies.
We find that applying the \GOLS{} algorithm with the \qnm{} frequencies corresponding to a different physical spin does not yield good fits.}.
\begin{figure}
\includegraphics[width=\factor\textwidth]{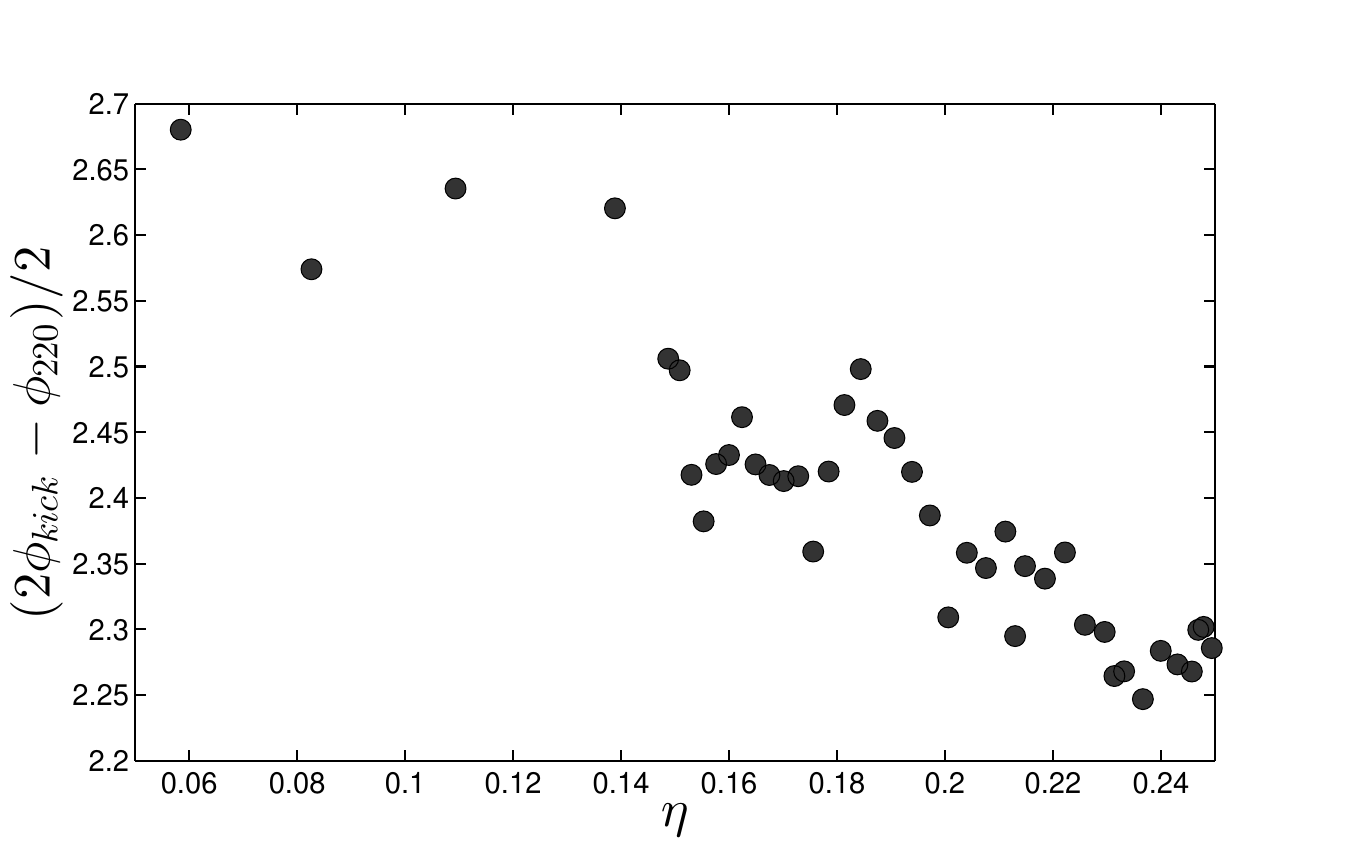}
\caption{Difference between phase of $(l,m,n)=(2,2,0)$ \qnm{} excitation ($10\,M$ after the peak luminosity in $\psi^{\,\mathrm{NR}}_{22}$) and the scaled kick direction, $m\phi_{kick}$ (Sec.\ref{sec:MAPPING}).}
\label{fig:phi0_22_sys}
\end{figure}
\par Even so, results for $A_{lmn}$ may be intrinsically biased if the data are not actually dominated by \qnm{s}.
This is the case if the fitting region is chosen either too close to the merger regime, or so far away that irregular numerical noise dominates.
For this reason, independent measures of the $|A_{lmn}|$'s consistency with perturbation theory are needed.
In this section we consider two such measures, and discuss the limitations of our results.
\subsection{Fitting region effects}
\label{sec:PT_Consistency:FittingRegion}
The first estimate of consistency is mentioned at the end of Sec.~\ref{sec:MULTI_METHOD}-\ref{note:error_bars}: the effect of \rd{} start time, $T_0$, on $A_{lmn}$.
Here we will discuss the effect of $T_0$ on $A_{lmn}$ from two perspectives.
\paragraph*{\textbf{Changing Scales.---}} \label{sec:Result_Limitations:Start_Time} On one hand, we may ask why defining $T_0$ relative to the peak luminosity of $\psi^{\mathrm{NR}}_{22}$ has been found to yield well-behaved maps between initial binary parameters and \qnm{} excitations.
For example, if one defines $T_0$ relative to the peak of  $\psi^{\mathrm{NR}}_{22}$ rather than its luminosity, then seemingly irregular oscillations are introduced into the dependence of each fundamental mode's $A_{lmn}$ on symmetric mass-ratio.
This suggests that there is something about the peak luminosity that serves as a consistent reference for how the system is evolving in the \rd{} regime.
This postulate is supported by our analysis of each $A_{lmn}$ phase in Sec.~\ref{sec:MAPPING}, where we found that when using the peak luminosity as a reference point, the complex phase of each $A_{lmn}$ was dependent on $m$ time the systems final kick direction with an offset of $m\phi_0$ that is largely independent of initial parameters (\eqn{\ref{eq:intrinsic_phase}}).
This means that the phase evolution of each \rd{} waveform, relative to the time of the peak luminosity, is approximate for the systems considered here.
In other words, the choice to measure time relative to the peak luminosity appears to be approximately the same as choosing $T_0$ such that $\phi_0$ is constant.
\par However, there is a discrepancy here: we have chosen $T_0 = 10 M$ in units of the system's Arnowitt-Deser-Misner (ADM) mass \cite{Kamaretsos:2012bs}, not the final black hole mass $M_f$, meaning that while the physical scale of the system($M_f$) changes, our reference length $T_0$ stays fixed.
This along with the dependence of each \qnm{} frequency on the final system state, $\{M_f, j_f\}$, should contribute to a systematically varying $\phi_0$.
The systematic dependence of $\phi_0$ is shown in \fig{\ref{fig:phi0_22_sys}} against $\eta$ ( $\eta$ is proportional to $j_f$).
\par As with choosing the peak of $\psi^{\mathrm{NR}}_{22}$ rather than its luminosity as a reference point, we might expect seemingly irregular oscillations to appear in the dependence of some $|A_{lmn}|(\eta)$.
In particular, while further study is needed, the above argument is a likely explanation for the fluctuations of some modes around $\eta=0.18$ (e.g. $|A_{320}|$ and $|A_{210}|$ in \fig{\ref{fig:fundamentals}}, and the modes in \fig{\ref{fig:overtones_and_2ndOrder}}).
\paragraph{\textbf{Different Start Times.---}} On the other hand, different fitting regions incur different amounts of numerical noise which may bias results.
Therefore we have chosen to quantify this measurement error by considering different fitting regions, and then rescaling our results to be relative to $T_0=10\,M$ after the peak in $\psi_{22}^{\,\mathrm{NR}}$'s luminosity.
This measure of consistency answers the question {\xquote{How much does the recovered \qnm{} behave like a damped sinusoid?}} and may be quantified by rescaling $A_{lmn}|_{T_0}$ according to its complex \qnm{} frequency
\begin{equation}
\label{eq:rescale_Almn}
A_{lmn}|_{T_0} \; \approx \; A_{lmn}|_{T_0'} \; e^{i\tilde{\omega}_{lmn}\;(T_0-T_0')} \;.
\end{equation}
In the ideal case, where the estimated $A_{lmn}$ behaves exactly as a decaying sinusoid from $T_0$ to $T_0'$, \eqn{\ref{eq:rescale_Almn}} becomes an equality.
This method was utilized to make the error bars throughout this paper.
\par While we find that the effects of choosing different $T_0$ are inherently systematic\footnote{To the left of \rd{} is the nonlinear merger, and to the right is numerical noise.}, they are also indicative of an optimal start of \rd{} that is generally about $10 M$ after the peak luminosity in $\psi^{\mathrm{NR}}_{22}$ (Appendix \ref{app:rd_start}); however, in some cases the effective \rd{} fitting may be performed up to $2 M$ after the peak luminosity.
An expanded description of fitting region effects is given in Appendix \ref{app:rd_start}.
\subsection{Inner-product ratios}
\par An additional consistency test may be performed by taking advantage of \eqn{\ref{eq:PSINR_PT_SUM}} for different $\psi^{\mathrm{NR}}_{lm}$ \cite{Kelly:2012nd}. Noting that any \qnm{} may be found within multiple $\psi^{\mathrm{NR}}_{lm}$ of the same $m$, it follows that the ratio of their mixing coefficients may be estimated from fitting results, and then compared to analytic calculations via \eqn{\ref{eq:YSPROD}}.
\par For example, in the case of $\psi^{\mathrm{NR}}_{33}$ and $\psi^{\mathrm{NR}}_{43}$, \eqn{\ref{eq:PSINR_PT_SUM}} gives that
\begin{align}
	\nonumber
	\psi^{\,\mathrm{NR}}_{33}(t) \, = \, A_{330} \, \sigma_{3330} \, e^{i\tilde{\omega}_{330}t} \, + \, ...
\end{align}
and
\begin{align}
	\nonumber
	\psi^{\,\mathrm{NR}}_{43}(t) \, = \, A_{330} \, \sigma_{4330} \, e^{i\tilde{\omega}_{330}t} \quad \quad \; \\
	\nonumber
	+ \, A_{430} \, \sigma_{4430} \, e^{i\tilde{\omega}_{430}t} \, + \, ...
\end{align}
By comparing terms, and recalling that the \GOLS{} algorithm gives a measure for terms in the above sum via \eqn{\ref{eq:MULTI_BETA}} $$\beta_{l'lmn}\;=\;A^{\,Est.}_{lmn}\,\sigma^{\,Est.}_{l'lmn}\,,$$ we see that the ratio, $\sigma_{l'lmn} / \sigma_{llmn}$ may be estimated directly from the results of multimode fitting.
For brevity, we shall limit our discussion to the fundamental modes.
For clarity, we will make a distinction between the \pt{} result derived from \eqn{\ref{eq:YSPROD}}
\begin{align}
	\label{eq:product_ratio_pt}
	\lambda^{\,\mathrm{PT}}_{l'm} \; = \; &\frac{\sigma_{l'lm0}}{\sigma_{llm0}} \; ,& \quad \;\;
\end{align}
and the multimode fitting estimate
\begin{align}
	\label{eq:product_ratio_nr}
	\lambda^{\,\mathrm{NR}}_{l'm} \; = \; &\frac{\beta_{l'lm0}}{\beta_{llm0}}& \\
	 = \; &\frac{ \sigma^{\,Est.}_{l'lm0}\,A^{\,Est.}_{lm0} }{ \sigma^{\,Est.}_{llm0}\,A^{\,Est.}_{lm0} } \;\;.&
	\nonumber
\end{align}
\par The three panels of \fig{\ref{fig:product_ratios}} compare $\lambda^{\,\mathrm{NR}}_{lm}$ to $\lambda^{\,\mathrm{PT}}_{lm}$ for $l=m=\{2,3,4\}$.
Because $\lambda^{\,\mathrm{NR}}_{lm}$ is insensitive to waveform phase, we have included results for three waveforms with lower symmetric mass-ratios.
\par While consistency between \pt{} and our numerical results is seen in all cases, our estimate $\lambda^{\,\mathrm{NR}}_{44}$ does systematically deviate from $\lambda^{\,\mathrm{PT}}_{lm}$ by roughly $10\%$ on $0.20 < \eta < 0.25$.
As suggested by our discussion in Sec. \ref{sec:SINGLE_ERR}, we consider this deviation to be the result of $|A_{440}|$ approaching the magnitude of numerical noise.
Moreover, this deviation was found to be exacerbated by the addition of mirror modes(Sec.~{\ref{sec:intro:NR_meets_PT}), the removal of the second order modes, or both.


\subsection{Limitations of results}
\label{sec:Result_Limitations}
\par While finite spatial and temporal \nr{} resolution limits the frequencies and multipoles that we are able to consider, we find that our results are stable with respect to the resolutions discussed in Sec. \ref{sec:SINGLE_ERR}.
This also suggests that gauge and near-field effects are not significantly manifested for the majority of our results\footnote{See \citep{Kelly:2012nd} for an expanded discussion.}.
However our consideration of the apparent second order modes carries a more basic limitation: we currently lack detailed knowledge about their structure.
Moreover, our lacking many simulations in the very unequal mass-ratio regime presents another limitation.
\paragraph*{\textbf{Second Order Modes.---}} As analytic calculations of second order Kerr \qnm{s} are lacking, there exists a tension in the existing literature.
\par On one hand, analytic studies such as that of Ioka and Nakano \cite{Ioka:2007ak} suggest that second order perturbations result in \qnm{s} proportional by products of two first order modes.
On the other hand, Pazos \textit{et al} \cite{Pazos:2010xf} found that, for spherically symmetric initial data, scalar wave scattering off of a Schwarzschild black hole results in second order excitations whose frequencies are \textit{the same} as those of first order modes.
\par In this study (Sec.~\ref{sec:Beyond}) we find second order excitations that appear to be largely driven by two first order \qnm{s}, with frequencies that are sums of two first order frequencies.
However, as our analysis approach has been designed to only extract spheroidal information post-merger, it cannot directly untangle mode coupling effects that would be consistent with \cite{Pazos:2010xf}.
Therefore, our findings may indeed be consistent with both \cite{Pazos:2010xf} and \cite{Ioka:2007ak}.
We expect that an analytic study, analogous to Leaver's work \cite{leaver85}, but for second order Kerr perturbations \cite{Campanelli:1998jv}, may elucidate the matter.
\par Among the subtleties that should be addressed, we expect the degeneracy of the sum and difference tone spectrum to play an important role: when considering the entire set of possible second order modes, one quickly finds exact or near degeneracies between \qnm{} frequencies with $l_1 \neq l_2$ and $m_1 \neq m_2$.
Here, the second order modes with the lowest $l=m$ indices, such as (2,2,0)(2,2,0) and (2,2,0)(3,3,0), are not only free from degeneracy at this level, but appear to be the most prominent.
\par On a more rudimentary note, we do caution that, for the apparent second order modes discussed in Sec.~\ref{sec:Beyond}, the overall proportionality constants (see Appendix \ref{app:Fit_Coeffs}) are surely biased by the numerical limitations discussed in this and previous sections.
\paragraph*{\textbf{Very unequal mass-ratios.---}} Lastly, in regards to our fits for \qnm{} excitation on symmetric mass-ratio, a more basic limitation is the inability to include many points in the very unequal mass-ratio regime ($\eta<0.15$).
Therefore, while the fits presented in Sec.~\ref{sec:MAPPING} have been constructed to adhere to the extreme mass-ratio limit, they are, conservatively, only valid within the presented range of $\eta$.
%
\section{Discussion of Results}
\label{sec:DISCUSS}
%
%
\begin{figure}[htb]
\includegraphics[width=\factor\textwidth]{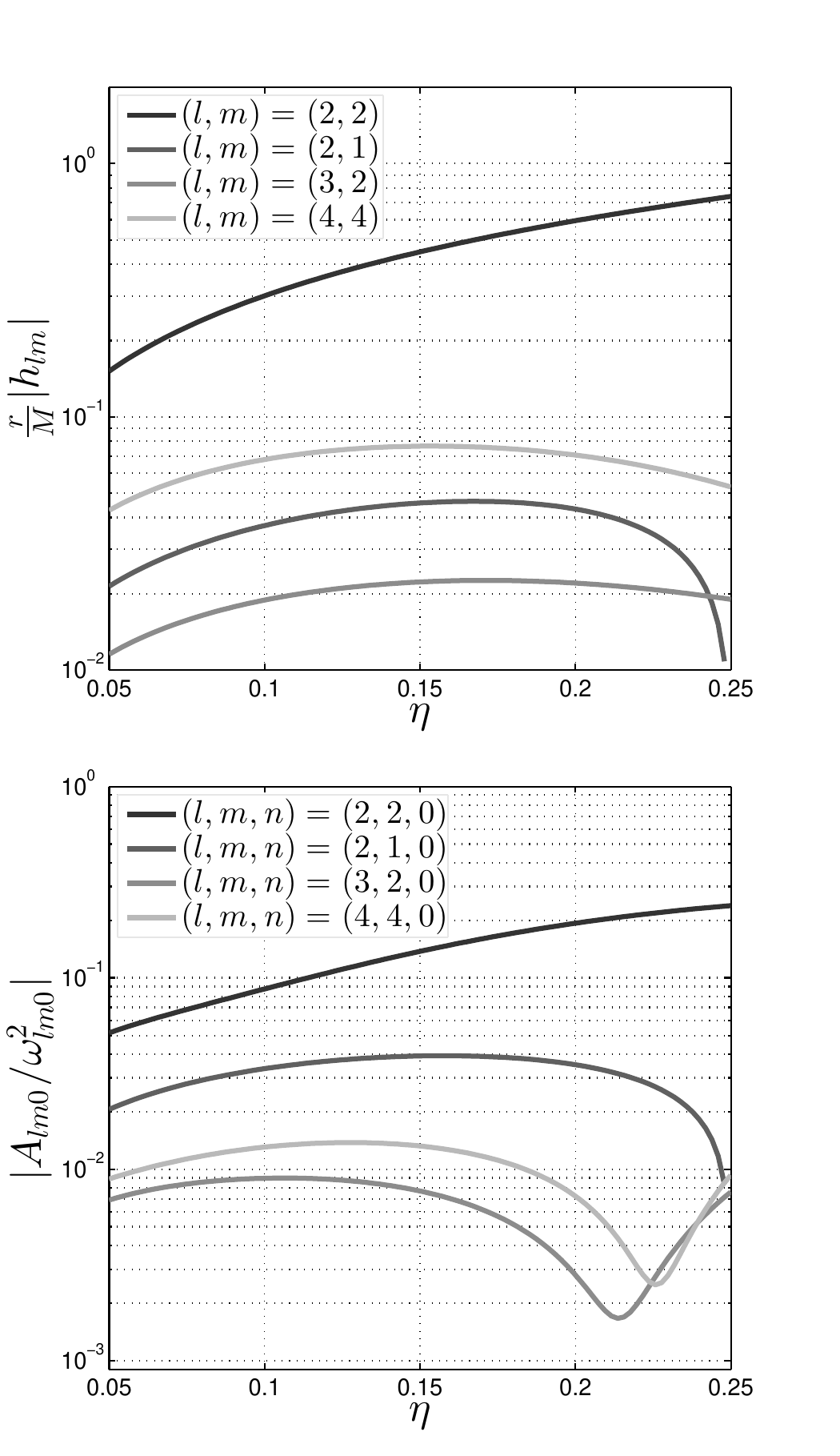}
\caption{Comparison of the PN strain amplitudes with \qnm{} amplitudes. {Top:} Amplitude of dimensionless Post-Newtonian strain for a selection of $(\ell,m)$ spherical multipoles. Values were calculated at $M\omega=0.18$ using reference \cite{oai:arXiv.org:0802.1249}. {Bottom:} Amplitude only fits for fundamental \qnm{} excitations.  }
\label{fig:pn_amplitudes}
\end{figure}

\par In this section, we comment on the potential relevance of subdominant \qnm{s} to \rd{} templates and the relevance of our results to perturbation theory.
\subsection{Perturbation theory comments}
\label{sec:DISCUSS:PTCOMMENTS}
Pending an analytic description of \qnm{} excitation for initially nonspinning, quasicircular \bbh{} merger, akin to \cite{Zhang:2013ksa}, and a better understanding of the higher order Kerr spectrum, akin to \citep{Zimmerman:2014aha}, we have found that a \pn{}-like prescription effectively models \qnm{} excitation for the systems studied.
%
%
The success of this model suggests that a well-defined analytic description exists, and that its predictions may be directly compared to the fitting coefficients in \eqns{eq:fit_fun_A220}{eq:fit_fun_A550}.
When directly compared to its \pn{} counterparts, our model also illuminates the qualitative differences between the inspiral regime, where \pn{} is valid, and the postmerger \rd{} regime.
\begin{figure*}[t]
\begin{center}
\includegraphics[width=0.9\textwidth]{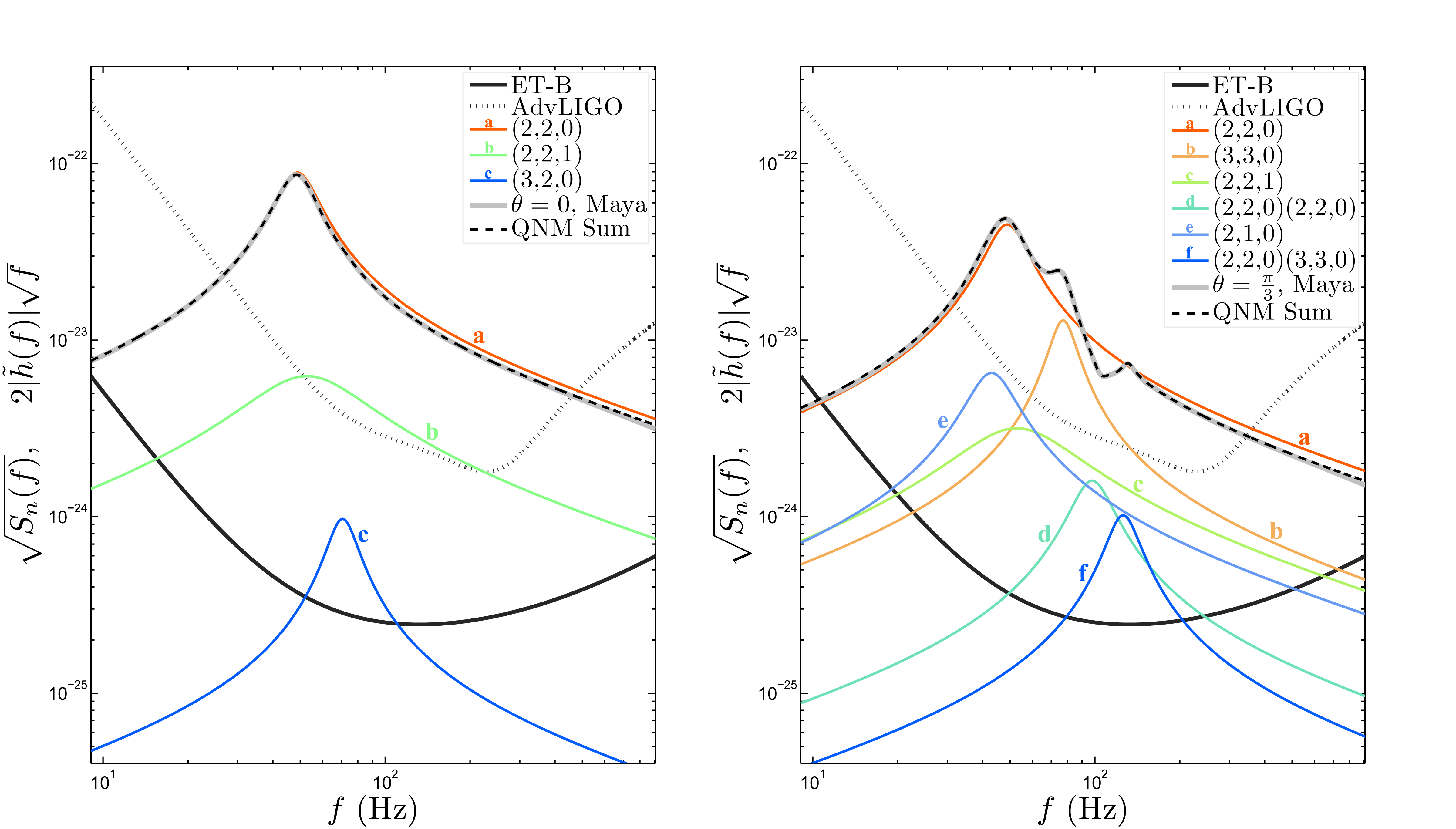}
\caption{Frequency domain envelopes of strain and fitted \qnm{} amplitudes for a 2:1 mass-ratio system ($\eta=0.22$) of 350 $M_\odot$, at a distance of 100 Mpc. Left: Signal for line of sight along final spin direction (e.g. $(\theta,\phi)=(0,0)$). Right: Line of sight $\pi/3\;rad$ with respect to final spin direction, $(\theta,\phi)=(\pi/3,0)$. Noise curves for the Einstein Telescope and Adv. LIGO are shown for reference. For each panel, the color of each quasinormal mode curve, along with its relative position, label the mode's contribution to total signal to noise ratio. In each case, the $(l,m,n)=(2,2,0)$ mode is the most dominant. }
\label{fig:hf_with_sensitivity}
\end{center}
\end{figure*}
\par In particular, \fig{\ref{fig:pn_amplitudes}} shows the qualitative differences between the spherical multipolar \gw{} emission predicted by \pn{} (top panel), and the fundamental spheroidal emission (bottom panel) presented here.
On one hand, similarities between the $(l,m,n)=\{(2,2,0),(2,1,0)\}$ \qnm{s} and their \pn{} counterparts may suggest that they are connected by a largely linear process.
On the other hand, the clear differences between \pn{} predictions, and the $(l,m,n)=(3,2,0)$ and $(4,4,0)$ \qnm{s} may suggest a region of nonlinear response between $\eta=0.1$ and $\eta=0.24$.
Further study is needed to precisely clarify whether or not this is the case.
\par Despite our current limited understanding of the underlying physics, the local minima seen in \fig{\ref{fig:pn_amplitudes}} suggest that the $(l,m,n)=(3,2,0)$ and $(4,4,0)$ \qnm{s} are less likely to be relevant for detection in the $\sim$2:1 mass-ratio ($\eta\approx 0.{22}$) regime. This point, in addition to our descriptions of the overtones and second order modes (Sec.~\ref{sec:MAPPING}), allows us to make qualitative comments on the relevance of \qnm{s} to template accuracy and mode detectability.
\subsection{Template comments}
\label{sec:DISCUSS:TEMPL}
\par While template accuracy and mode detectability are topics whose full treatment is beyond the current work, we are able to briefly comment on the impact of subdominant \qnm{s} on the \snr{} of \rd{} signals.
To do so, we will reconsider the 2:1 mass-ratio binary discussed in the introduction (\fig{\ref{fig:full_ringdown}}).
\par Specifically, let us contemplate an idealized scenario where a \rd{}-only template is being used to search for a potential signal as observed by either the \et{}, or \aligo{}.
For simplicity we will assume that either detector is equally sensitive over the solid angle, and that there are no glitches in detector sensitivities as presented in Refs.~\cite{Ajith:2009fz,Mishra:2010tp}.
To completely constrain our example, we will consider only templates made with binary parameters identical to that of the signal: final mass 350 $M_\odot$, at a distance of 100 Mpc, initially nonspinning, $\eta=0.22$, and quasicircular.
We are only interested in the effect of subdominant \qnm{s} on the estimated \snr{}.
\par If the signal, $\tilde{s}(f)$, is the frequency domain counterpart of \fig{\ref{fig:full_ringdown}}'s waveform, and the  template, $\tilde{h}(f)$, is composed of some superposition of \qnm{s} according to the Fourier transform of \eqn{\ref{eq:PSI4PT}}, then the \snr{} is given by
\begin{align}
	\label{eq:snr}
	\rho \; = \; \frac{(\tilde{s}(f)|\tilde{h}(f))}{\sqrt{(\tilde{h}(f)|\tilde{h}(f))}}
\end{align}
where
\begin{align}
	(a(f)|b(f)) \; \equiv \; 2 \, \int_{-\infty}^\infty \! \frac{a^*(f)\,b(f)}{S_n(f)}  \mathrm{d}f
\end{align}
and $S_n$ is the power spectral density (PSD) of the detected noise \cite{Pekowsky:2013ska,Ajith:2009fz,Mishra:2010tp}.
\par In the best case scenario, where the signal and template are identical, $\rho$ takes on its maximal value, $\rho_{max}$. Table~\ref{tab:total_snr} lists the values of $\rho_{max}$ for the orientations shown in \fig{\ref{fig:full_ringdown}}.
\begin{table}[H]
\caption{\label{tab:total_snr} Maximal \snr{} values, $\rho_{max}$, for \et{} and Advanced LIGO (Adv. LIGO)  detectors at two different orientations with respect to the final \bh{'s} spin direction: $(\theta,\phi)=\{(0,0),(\pi/3,0)\}$. Final mass 350 $M_\odot$, distance 100 Mpc, initially nonspinning, $\eta=0.\overline{22}$, quasicircular.}
\begin{center}
\begin{tabular}{|c||c|c|}
	\hline
	\hline
	$(\theta,\phi)$ & \multicolumn{2}{|c|}{$\rho_{max}$} \\ \cline{2-3}
	 & \aligo{} & \et{} \\
	\hline
	$(0,0)$ & 10.58 & 160.79
	\\ \hline
	$(\pi/3,0)$ & 6.20 & 94.29 \\
	\hline
	\hline
\end{tabular}
\end{center}
\end{table}
\par We now ask which \qnm{s} contribute the most to the total \snr{} for each of the cases above.
To answer this question, we sequentially determine which $N$-mode template recovers the largest percent of $\rho_{max}$.
For example, if we denote the recovered \snr{} of each $N$-mode template to be $\rho_*$, then in the case of \aligo{}, the 1-mode template that recovers the largest percentage of $\rho_{max}$ contains only the $(l,m,n)=(2,2,0)$ \qnm{}.
This is the case for $\theta=0$, where $\rho_*=0.9986\rho_{max}$, and for $\theta=\pi/3$, where $\rho_*=0.9749\rho_{max}$.
If we ask which additional \qnm{} results in the largest $\rho_*$ at $\theta=\pi/3$, then $(3,3,0)$ proves to be the next most important, with $\rho_*=0.9837\rho_{max}$.
Taking another step forward, we find that the best 3-mode template for \aligo{} at $\theta=\pi/3$ includes the $(2,2,0)$, $(3,3,0)$ and $(2,2,1)$ \qnm{s}, with an \snr{} of $\rho_*=0.9902\rho_{max}$.
\begin{table}[H]
\caption{\label{tab:hf_snr_values} Recovered \qnm{s} and  estimated fractional SNR values for Advanced LIGO (Adv. LIGO) and the Einstein Telescope. Under each detector heading, values for the SNR found using only one mode, $\rho_1$, and values for using many modes, $\rho_*$, are shown. In the case of $\rho_*$, the number of \qnm{s} used in the template increases from top to bottom. This may be seen in the first row of each case, where $\rho_*=\rho_1$.}
\begin{tabular}{|l||c|| p{1cm} | p{1cm} ||p{1cm}|p{1cm}|} 
  \hline 
  \hline 
  $(\theta,\phi)$ & Mode & \multicolumn{2}{|c||}{Adv. LIGO} & \multicolumn{2}{c|}{\et{}} \\ \cline{3-6} 
  & $(l,m,n)$ &  $\rho_1\;(\%)$ & $\rho_{\ast}\;(\%)$ & $\rho_1\;(\%)$ & $\rho_{\ast}\;(\%)$ \\ 
  \hline 
 
\multirow{3}{*}{$(0,0)$} &  
$(2,2,0)$ & $99.865$ & $99.865$ & $99.880$ & $99.880$\\ 
& $(2,2,1)$ & $89.461$ & $99.986$ & $86.956$ & $99.989$\\ 
& $(3,2,0)$ & $62.561$ & $99.997$ & $59.026$ & $99.998$\\ 
 
\hline 
\multirow{6}{*}{$(\frac{\pi}{3},0)$} &  
$(2,2,0)$ & $97.494$ & $97.494$ & $98.348$ & $98.348$\\ 
& $(3,3,0)$ & $63.946$ & $98.365$ & $60.932$ & $98.801$\\ 
& $(2,2,1)$ & $86.457$ & $99.023$ & $85.537$ & $99.349$\\ 
& $(2,1,0)$ & $41.464$ & $99.558$ & $92.670$ & $99.685$\\ 
& $(2,2,0)(2,2,0)$ & $92.069$ & $99.795$ & $40.896$ & $99.886$\\ 
& $(2,2,0)(3,3,0)$ & $30.870$ & $99.934$ & $27.192$ & $99.957$\\ 
 
  \hline 
  \hline 
\end{tabular} 

\end{table}
Table~\ref{tab:hf_snr_values} lists the percentages of $\rho_{max}$ recovered up to the 6-mode template for $\theta=\pi/3$ and up to the 3-mode template for $\theta=0$.
Figure~\ref{fig:hf_with_sensitivity} is a graphical representation of Table~\ref{tab:hf_snr_values}, and displays each frequency domain \qnm{} against the \et{} and \aligo{} PSDs.
\par This simple numerical experiment suggests that the greater the angle between the detector's line of sight and the \bh{'s} final spin direction, the more \qnm{} information is needed to model the signal up to $99\%$ of $\rho_{max}$.
While the orientation dependence and impact of multipoles with $l>2$ on detectability is a topic of active interest \cite{Healy:2013jza,Pekowsky:2012sr,Capano:2013raa}, and previous studies of adding fundamental \qnm{s} of $\ell>2$ to \rd{}-only templates have suggested a significant effect on event loss \cite{Berti:2007zu,Caudill:2011kv}, our example demonstrates that the $(l,m,n)=(2,2,1)$ overtone may play a meaningful role.
%
%
Further study, similar to \cite{Berti:2007zu}, is needed to better quantify its significance.
\par Intriguingly, although Table~\ref{tab:hf_snr_values} shows that the second order \qnm{s} may only add a minuscule amount to the total \snr{}, their contribution to the frequency domain features in \fig{\ref{fig:hf_with_sensitivity}} raises the possibility of their being identified postdetection.
\par Finally, in light of the \qnm{} amplitude and phase results presented in Sec.~\ref{sec:MULTI_FITs}, our toy example also allows us to consider what information about the remnant \bh{} may be learned.
It is well known that the scaling of \qnm{} frequencies with remnant mass means that the detection of at least two \qnm{} frequencies is required to estimate the final mass and spin of the system \cite{Berti:2009kk,Berti:2007zu,Berti:2005ys,Kamaretsos:2012bs,Gossan:2011ha}.
This information, along with the relative amplitudes may also yield information about the initial binary, and perhaps even final spin orientation \cite{Kamaretsos:2011aa,Kamaretsos:2012bs}.
Of the current study, if two \qnm{} frequencies are detected, allowing for the identification of each frequency's $(l,m,n)$, then a rearrangement of \eqn{\ref{eq:phi0_definition}} suggests that information about the recoil angle relative to the line of sight may also be estimated via
\begin{align}
	\label{eq:measurable_kick_angle}
	\phi_{kick} \; \approx \; \,    \frac{\phi_{220}}{2} - \phi_0 \;.
\end{align}
\par The applicability of this potential measure is the subject of a future study.
%
\section{Conclusion}
%
\par Our in-depth analysis of \nr{} entrance into \rd{} has provided us with a wealth of information about the excitation of \qnm{s}.
We have found evidence for nonfundamental spheroidal \qnm{} excitations within the residuals of single-mode \qnm{} fits  (Sec.~\ref{sec:SINGLE_RES}).
By developing a method to estimate these spheroidal components (Sec.~\ref{sec:MULTI_METHOD}), we have presented a review of \qnm{} excitations including and beyond the fundamentals, and we have discovered that the phase of these excitations is affected by the remnant \bh{'s} final kick direction (Sec.~\ref{sec:Almn_Fitting}).
\par \qnm{} excitations are well modeled by a \pn{}-like expansion (Sec.~\ref{sec:MAPPING}), and that our estimates for the excitation amplitudes are largely consistent with perturbation theory, within the limits of knowledge and numerical accuracy available at the time of this study (Sec.~\ref{sec:PT_Consistency}).
\par To make our results available for the construction of \rd{} related \gw{} templates, we have tabulated related fitting coefficients in Appendix~\ref{app:Fit_Coeffs}.
\par We studied the relevance of our results for \gw{} detection with the \rd{} of a 2:1 mass-ratio system of initially nonspinning \bh{s}.
For this case, we find that the $l=m=2$, $n=1$ overtone is the most dominant, and that that it is the second most significant \qnm{} when the remnant \bh{} is observed along its final spin axis (\fig{\ref{fig:hf_with_sensitivity}} left panel).
This case also demonstrates that the apparent $l=m=2$ second mode, while minuscule in comparison to its first counterpart, may be more significant than higher $l$ \qnm{s} at similar frequencies (\fig{\ref{fig:hf_with_sensitivity}} right panel).
Moreover, this case is consistent with the expectation that as the line of sight deviates from the final \bh{} spin direction, more \qnm{s} are needed to accurately represent the signal (Table~\ref{tab:hf_snr_values}).
\par But as informative as our example 2:1 mass-ratio system may be, its shortcoming are clear.
It demonstrates that when modeling \rd{} the $(l,m,n)=(2,2,1)$ can play a role comparable to that of the higher fundamental \qnm{s} (Table \ref{tab:hf_snr_values}), but to solidify this statement, and it's relevance to high mass templates, a full orientation study is needed.
We have also seen that apparent second order \qnm{s} might contribute to \rd{'s} frequency domain features (\fig{\ref{fig:hf_with_sensitivity}}), but the full extent to which these modes are relevant cannot be assessed without more accurate \nr{} simulation, and a better understanding of the second order structure of Kerr perturbations.
Intriguingly, we have also seen that \qnm{} phase carries information of how the remnant \bh{} is oriented relative to its recoil velocity.
While our example system demonstrates that this might allow for an estimation of the recoil direction relative to the line of sight, the scope of the estimation as presented here is only a first step.
We look forward to the exploration of this possibility in future work.
%
\section{Acknowledgments}
We are very grateful to Kostas Kokkotas, Pablo Laguna, Vitor Cardoso, Emanuele Berti, Sam Finn and Larne Pekowsky for their helpful input and discussion. We also gratefully acknowledge to support of NSF Grants No. 0955825 and No. 1212433. Numerical simulations were carried out at Teragrid  TG-PHY120016 and on the CRA Cygnus clusters.
\appendix

\section{Fitting Coefficients for \qnm{} Excitations}
\label{app:Fit_Coeffs}
For convenience, here we have collected all fitting formulas and related coefficients.
In particular, if one is interested in the \qnm{} excitations from initially nonspinning, quasicircular binary black hole coalescence, then we present the following algorithmic description to apply the model presented in Sec.~\ref{sec:MAPPING}.
For additional convenience, a basic usage and plotting example has been made available in reference \cite{London:2016err_web}.
\par The primary inputs of our model are the binary's component masses, $\mathrm{m_1}$ and $\mathrm{m_2}$.
The primary output of our model is the \rd{} portion $\psi_4(t)$, starting 10 ($M $) after the peak luminosity in the $l=m=2$ spherical multipole.
Therefore, throughout what follows, $t=0$ corresponds to 10 ($M $) after the $l=m=2$ spherical multipole, and values of $t<0$ are to generally be considered outside of the fit's domain of applicability.
%
%
\par First, given $\mathrm{m_1}$ and $\mathrm{m_2}$, one may calculate the symmetric mass-ratio via $$\eta = \frac{\mathrm{m_1}\mathrm{m_2}}{\mathrm{m_1}+\mathrm{m_2}}\,.$$
With the symmetric mass-ratio, one may use a phenomenological fitting formula to quickly estimate the remnant \bh{'s} final mass, $M$, and dimensionless spin, $j=S/M^2$.
While we present fitting formulas in Appendix~\ref{app:mf_jf}, an alternative formula may be found in \cite{Rezzolla:2007rd}.
\par Now with the final \bh{'s} parameters at hand, individual \qnm{} frequencies, $$\tilde{\omega}_{lmn}=\omega_{lmn}+i/\tau_{lmn}\;,$$ may be most readily obtained by using the fitting formulas presented in \cite{Berti:2005ys}.
Alternatively one may use the tabulated values for $M\omega_{lmn}$ available at \cite{Berti:2005ys_web}.
\par We have that estimates for the complex \qnm{} excitation factors, $A_{lmn}$, may be found by evaluating the following series of equations:
\begin{align}
	\delta_m(\mathrm{m_1},\mathrm{m_2}) \, \equiv \,  &\frac{|\mathrm{m_1} + (-1)^m\mathrm{m_2}|}{\mathrm{m_1} + \mathrm{m_2}}&
\end{align}
\begin{align}
	A_{lmn} \, = \;  &\tilde{\omega}_{nlm}^2 \; \delta_m(\mathrm{m_1},\mathrm{m_2}) \;\eta^{1+n} \, \sum_{u=0} \, |a_u|e^{i\alpha_u} \, \eta^u \,&
\end{align}
Values for $|a_u|$ and $\alpha_u$ are given in \eqns{eq:fit_fun_A220}{eq:fit_fun_A550}.
\par For the second order \qnm{s} discussed in Sec. \ref{sec:Beyond}, we have that $$A_{(l_1,m_1,n_1)(l_2,m_2,n_2)} = \mu_{(l_1,m_1,n_1)(l_2,m_2,n_2)}A_{l_1,m_1,n_1}A_{l_2,m_2,n_2}\,,$$ where for the (2,2,0)(2,2,0) mode we find that $$\mu_{(2,2,0)(2,2,0)}=5.3956\,,$$ and for the (2,2,0)(3,3,0) mode, $$\mu_{(2,2,0)(3,3,0)}=4.6354\,.$$
%
%
Keeping in mind that all tabulated coefficients correspond to $T_0=10$ ($M $), the full time domain \rd{} waveform may be calculated by first evaluating the spheroidal harmonics, $_{-2}S_{lm}(j\tilde{\omega}_{lmn},\theta,\phi)$ (via \cite{leaver85}), then evaluating
\begin{eqnarray}
    \nonumber
	\psi_4(t,\theta,\phi) = \; \frac{1}{r} \sum_{l,m,n} \, \psi_{lmn}^{\,\mathrm{PT}}(t) \, [_{-2}S_{lm}(j\tilde{\omega}_{lmn},\theta,\phi)]
\end{eqnarray}
where
\begin{eqnarray}
    \nonumber
	\psi_{lmn}^{\,\mathrm{PT}}(t) = \, A_{lmn} \, e^{i\tilde{\omega}_{lmn}t} \, .
\end{eqnarray}
Alternatively, one may calculate the spherical multipole moments by evaluating
\begin{eqnarray}
    \nonumber
	\psi^{\,\mathrm{NR}}_{l'm}(t) \, = \, \sum_{n,l} \, A_{lmn} \, \sigma_{l'lmn} \, e^{i\tilde{\omega}_{lmn}t}
\end{eqnarray}
where
\begin{eqnarray}
    \nonumber
	\sigma_{l'lmn} \, \equiv \, \int_{\Omega} \, _{-2}S_{lm}(j\tilde{\omega}_{lmn},\theta,\phi) \, _{-2}\bar{Y}_{l'm}(\theta,\phi) \, \mathrm{d}\Omega \,.
\end{eqnarray}
While we have suppressed the second order notation for simplicity, one may again impose the notion that each full second order \qnm{} corresponds to products of two first order modes.
With the two expressions for $\psi^{\,\mathrm{NR}}_{l'm}(t)$ and $\psi_{lmn}^{\,\mathrm{PT}}(t)$ above, we have completed our algorithmic description for calculating \rd{} waveforms using the initial binary's component masses.
\par While our discussion thus far has been limited to first and fundamental overtones, $n=0$ and $n=1$, it should also be noted that consistent evidence for the $n=2$, $l=m=2$, overtone may be readily observed by considering fitting regions closer to the $\psi^{\mathrm{NR}}_{lm}$ luminosity. Figure \ref{fig:second_22_overtone} displays this overtone scaled relative to $T_0=10$ ($M $). Though the general trend is reminiscent of the $n=0$ and $n=1$ overtones, the $n=2$ mode's faster decay rate corresponds to larger variation with fitting region (e.g. larger error bars).
\begin{figure}[htb]
\includegraphics[width=\factor\textwidth]{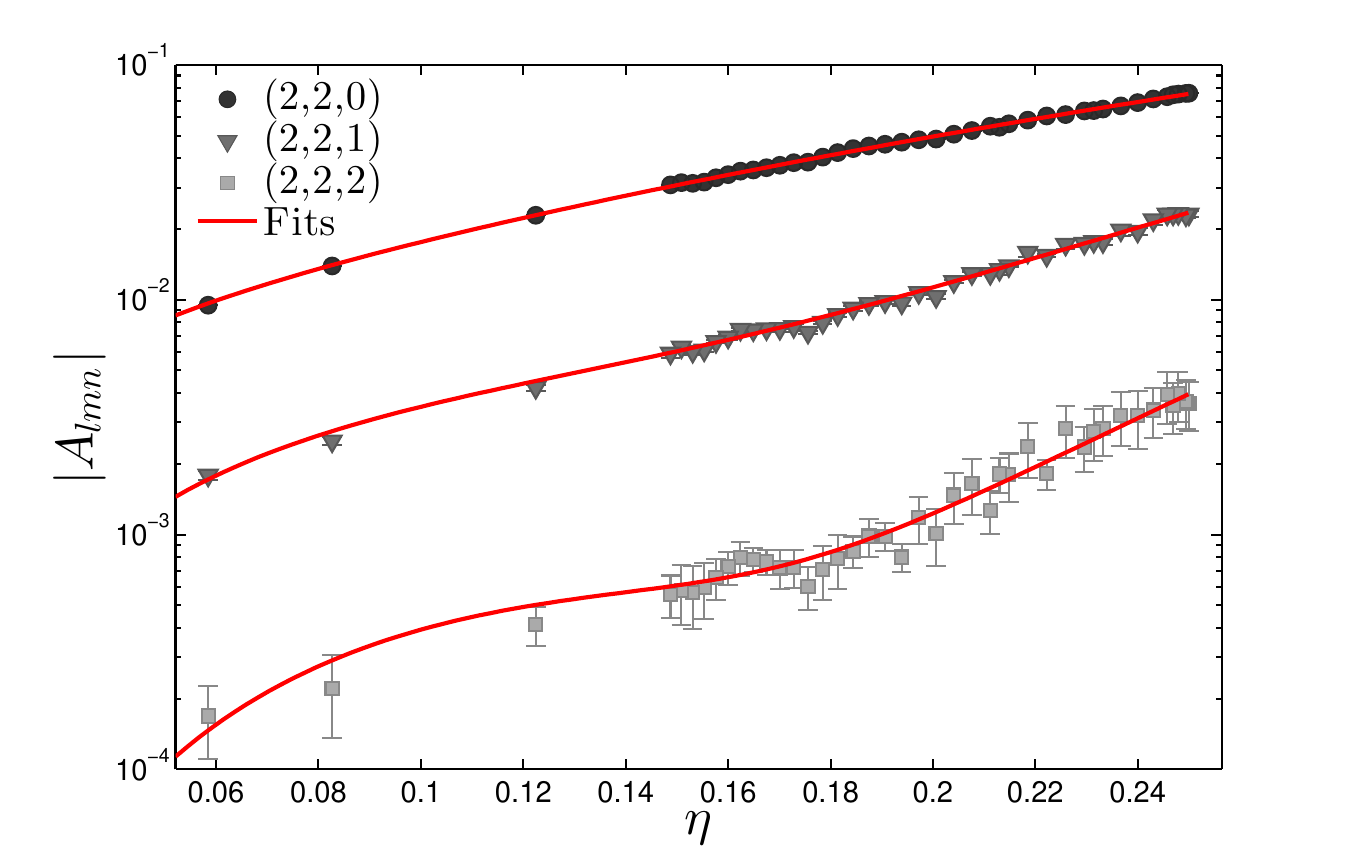}
\caption{\label{fig:second_22_overtone} The $n=0,1$ and $2$ overtones of the $l=m=2$ \qnm{} excitation recovered from \nr{} \rd{} for initially nonspinning unequal mass-ratio \bh{} binaries. The error bars were calculated as described in Sec.~\ref{sec:MULTI_METHOD}- \ref{note:error_bars}. }
\end{figure}
%
\section{The Start of Ringdown}
\label{app:rd_start}
While it is not possible to define an absolute start of \rd{}, we may make a practical definition by asking which potential \rd{} region is best modeled by \qnm{s} only.
\begin{figure}[htb]
\begin{center}
\includegraphics[width=0.5\textwidth]{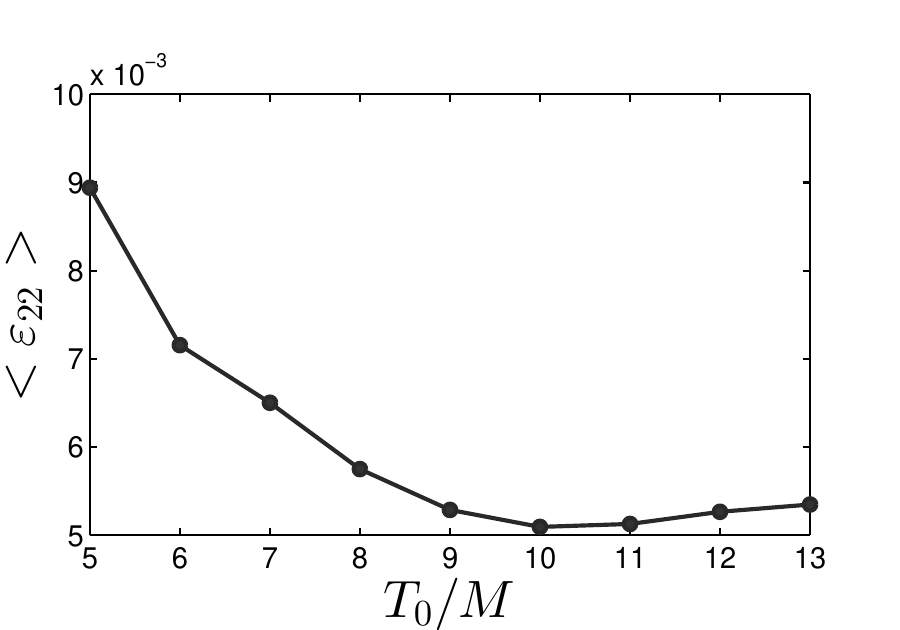}
\caption{Mean fractional root-mean-square error (\eqn{\ref{eq:RMSE}}) for the $l=m=2$ multipole with respect to the fitting region start time, $T_0$. Here the \GOLS{} (Sec.~\ref{sec:MULTI_METHOD}) algorithm was used to used to perform a multimode fit for each fitting region.}
\label{fig:best_T0}
\end{center}
\end{figure}
This question may be addressed by finding a local minimum in residual error with respect to fitting region start time.
To this end let us consider the multipole which is least effected by numerical errors: $\psi^{\,\mathrm{NR}}_{22}$.
Figure~\ref{fig:best_T0} shows its residual error [\eqn{\ref{eq:RMSE}}] on symmetric mass-ratio.
The trend observed here is inherently systematic as, when moving towards the peak in radiation, the data are no longer dominated by \qnm{s}, while, when moving away from the peak, numerical noise eventually dominates.
\par Consequently,  although there is a visible minimum at $T_0=10$ $(M )$, it is not the global minimum, as $\varepsilon_{22}$ fluctuates in the numerical noise following $T_0=13$ $(M )$. However, $10$ $(M )$ nevertheless gives us a practical starting point within which the majority of $\psi^{\,\mathrm{NR}}_{lm}$ is above the numerical noise floor.
%
\begin{table}[thb]
\caption{\label{tab:masspin_fit_coeffs} Fitting coefficients for $M_f(\eta)$ (\eqn{\ref{eq:mass_model}}) and $j_f(\eta)$ (\eqn{\ref{eq:spin_model}}).}
\begin{tabular}{|c|| c | c | c | c | c | c |} 
	\hline 
	\hline 
	          & $t_0$ & $t_1$ & $t_2$ & $t_3$ & $t_4$ & $t_5$ \\ 
	\hline 
	\hline 
	$M$ & $1$ & $-0.046297$ & $-0.71006$ & $1.5028$ & $-4.0124$ & $-0.28448$\\ 
	\hline 
	$j$ & $0$ & $3.4339$ & $-3.7988$ & $5.7733$ & $-6.378$ & $0$\\ 
	\hline 
	\hline 
\end{tabular} 

\end{table}
%
%
\section{Final Mass and Spin}
\label{app:mf_jf}
As noted in \cite{Rezzolla:2007rd}, the final mass and spin dependence on initial binary symmetric mass-ratio may be well fit by a polynomial in $\eta$.
Alternatively, the more recent study, Ref. \cite{Healy:2014yta}, shows that the final \bh{} parameters may also be well modeled as a power series in $\mathrm{m_1-m_2}$.
Here, we present a methodologically different fit than that presented in \cite{Rezzolla:2007rd} and \cite{Healy:2014yta}, while maintaining the $\eta$ parametrization of \cite{Rezzolla:2007rd}.
Specifically, when fitting final dimensionless spin, $j_f$, we choose to directly impose the boundary condition that as $\eta \rightarrow 0$, $j_f \rightarrow 0$. In particular, we fit
\begin{align}
\label{eq:spin_model}
	j_f(\eta) = \eta \, \sum_{k=1} \, t_{k} \, \eta^{k-1}
\end{align}
Similarly, when fitting final mass, $M_f$, we choose to directly impose the boundary condition that as $\eta \rightarrow 0$, $M_f\rightarrow 1$.
In particular, we fit
\begin{align}
\label{eq:mass_model}
	M_f(\eta) = 1 - \eta \, \sum_{k=1} \, t'_{k} \, \eta^{k-1}.
\end{align}
The fitting result for $j_f(\eta)$ is shown in \fig{\ref{fig:jf_on_massratio}}.
Fitting coefficients are tabulated in Table \ref{tab:masspin_fit_coeffs}.
While the fitting results here are consistent with \cite{Rezzolla:2007rd} and \cite{Healy:2014yta} within their fit's domain of applicability (deviations are within 1\% of the values reported), we expect that the forms given in \ref{eq:spin_model} and \ref{eq:mass_model} bias the fit towards the physically correct solution outside of the fitting domain.
\paragraph*{\textbf{Consistency with multimode Fit}.---} The numerical values used to make the above fits (Table \ref{tab:masspin_fit_coeffs}) were calculated using the isolated horizon formalism \cite{Frauendiener:2011zz}.
However, final \bh{} mass and spin may also be estimated using \rd{} fitting (e.g. \cite{Hannam:2010ec,Berti:2007zu}).
For the numerical runs considered here, we find that single mode fitting recovers the horizon estimate to within $\sim5\%$, while multimode fitting recovers the horizon estimate to within $\sim 0.5\%$.
This level of agreement is within the numerical error of the isolated horizon estimate.
%
\bibliographystyle{ieeetr}
\bibliography{refs.bib}


\end{document}